\documentclass[twocolumn,aps,pre,amsmath,amssymb,groupedaddress]{revtex4-2}
\usepackage{amsmath}
\usepackage{amssymb}
\usepackage{graphicx}
\usepackage{hyperref}
\makeatletter
\usepackage{units}

\usepackage{verbatim}
\usepackage{color}

\usepackage{dcolumn}
\usepackage{bm}
\usepackage{amsfonts}
\usepackage{pifont}
\usepackage{amsfonts}
\DeclareMathOperator{\sech}{sech}
\usepackage[bottom]{footmisc}

\newcommand{\beq}{\begin{eqnarray}}
\newcommand{\eeq}{\end{eqnarray}}

\makeatother
\newcommand{\bs}[1]{\boldsymbol{#1}}
\begin{document}
\title{Coarse-graining the vertex model and its response to shear}
\author{Gloria Triguero-Platero}
\affiliation{Institute for Theoretical Physics, Heidelberg University, D-69120 Heidelberg, Germany}
\author{Falko Ziebert}
\affiliation{Institute for Theoretical Physics, Heidelberg University, D-69120 Heidelberg, Germany}
\author{Luis L. Bonilla$^{*}$}
\affiliation{Department of Mathematics, Universidad Carlos III de Madrid, 28911 Legan\'es, Spain}
\affiliation{G. Mill\'an Institute for Fluid Dynamics, Nanoscience and Industrial Mathematics, Universidad Carlos III de Madrid, 28911 Legan\'es, Spain \\
$^*$Corresponding author. E-mail: bonilla@ing.uc3m.es
}

\date{\today}
\begin{abstract}
Tissue dynamics and collective cell motion are crucial biological processes. Their biological machinery is mostly known, and simulation models such as the \textit{active vertex model} (AVM) exist and yield reasonable agreement with experimental observations like tissue fluidization or fingering. However, a good and well-founded  continuum description for tissues remains to be developed. In this work we derive a macroscopic description for a two-dimensional cell monolayer by coarse-graining the vertex model through the Poisson bracket approach. We obtain equations for cell density, velocity and the cellular shape tensor. We then study the homogeneous steady states, their stability (which coincides with thermodynamic stability), and especially their behavior under an externally applied shear. Our results contribute to elucidate the interplay between flow and cellular shape. The obtained macroscopic equations present a good starting point for adding cell motion, morphogenetic and other biologically relevant processes. 
\end{abstract}

\maketitle

\section{Introduction}
\label{intro}

From the physical point of view, a tissue is a very complex material, since its constituents are active objects consuming energy and exerting forces onto the outside and between each other \cite{gil18,hak17,tre18,gia18,ale20}. Individual cells are assembled into tissues by coupling to their neighbors through specific transmembrane protein complexes (cadherins), which build cell junctions. The latter physically link the actomyosin cortices of neighboring cells, enabling force transmission between them \cite{gil18}. Many individual cellular processes (changes of cell shapes, cellular divisions, rearrangements, and extrusions) cause large-scale deformations of tissues. In the last decades, extensive research has been devoted to understand the links between cellular processes, tissue deformations and cohesive coordinated cellular motion \cite{may16,str17,lad17,xi19,ale20,loeber15} responsible for wound healing assays \cite{hak17,gia18,pou07}, cancer progression \cite{abe62,fri09,ben12,str20,moi19,bon20} and morphogenesis \cite{gil18,huf07,lec11,goo21}. A good macroscopic description of the mechanics and dynamics of tissue remains a major challenge
at the interface of physics and biology, although there have been many proposals, see Refs.~\cite{ale20,mar13,ish17,mer17,her21,tlili15}. 

The main objective of this work is to derive a macroscopic description for a two-dimensional flat tissue such as an epithelial monolayer
by a well-defined coarse-graining procedure. Successful 
and currently often-used
mesoscopic models describing tissues as a network of cells that fill space with no gaps between cells exist:
these are vertex and Voronoi models \cite{hon04,far07,huf07,fletcher14,alt17}, first used to describe the physics of foams \cite{gra00,wea84}. 
Interestingly, such models predict a jamming-unjamming (solid-liquid) transition at a critical mean shape index, which is the ratio between the mean cell perimeter and the square root of the mean cell area \cite{bi14,bi15}, and this has been observed in experiments \cite{mal17}. Here, we use the Poisson bracket method
\cite{maz06,cha95}
to coarse-grain the dynamics
governed by the free energy of the vertex model. 
We obtain macroscopic -- hydrodynamic -- equations, that keep track of the underlying cellular structure
due to the coupling to an equation for the average cellular shape tensor. 
We study the stability of the homogeneous phases,
reflecting the above-mentioned 
transition, as well as the effects of externally shearing the layer.

In Section \ref{sec:2}, we briefly discuss the widely used \textit{Active Vertex Model (AVM)} \cite{bar17}. It describes tissues as a network of polygonal cells forming a Voronoi tiling of the plane. In turn, the dual Delaunay triangulation of the plane uses the centers of the cells 
which underlie a dynamics governed by the vertex free energy function, 
as well as possibly additional active terms, typically intended to model cell motion.
Note that in the following
we treat only the passive version.
Usually overdamped dynamics is used in simulations \cite{bar17}, but 
underdamped dynamics with collective inertia has been recently proposed, allowing to capture more qualitative features of confluent cellular motion \cite{bon20} seen in experiments \cite{lv20,val20}.

The coarse graining procedure is then reviewed in Section \ref{sec:3}, following largely Ref.~\cite{her21}. In Section \ref{sec:4}, we give the average free energy density of the homogeneous phases in terms of the shape tensor that is then decomposed in trace, anisotropy and nematic order-like tensor fields
and discuss differences to previous works \cite{her21}.
The thermodynamic stability of the homogeneous phases is analyzed in Section \ref{sec:5}, where we show that the solid-liquid transition at the critical shape index is a pitchfork bifurcation from isotropic to anisotropic phases at a critical value of the line tension. Section \ref{sec:6} explains our choice of the kinetic coefficients and derives the resulting final continuum equations. The homogeneous phases according to these equations are studied in Section \ref{sec:7}, and Section \ref{sec:8} investigates the behavior under shear flow, resulting in an imperfect pitchfork bifurcation. Lastly, Section \ref{sec:9} contains our conclusions.

\section{Active vertex model}\label{sec:2}
\label{AVMsec}

\begin{figure}[t!]
\centering
    \includegraphics[width=\linewidth]{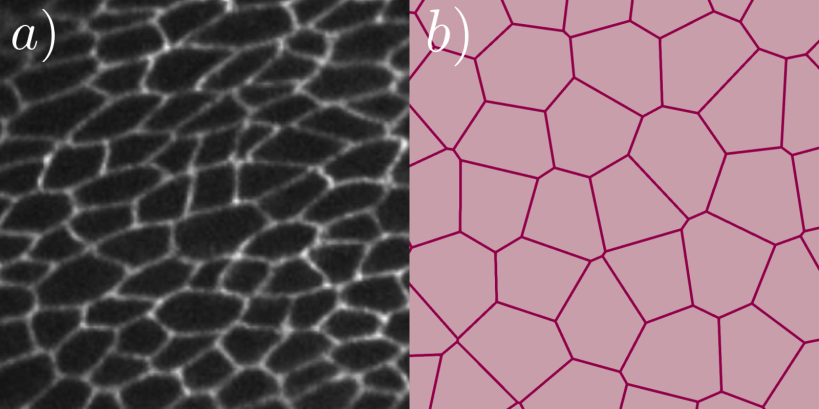}
\caption{a) Apical view of the wing imaginal disc epithelium in the Drosophila embryo (modified from \cite{hir17}). b) Snapshot of a  two dimensional cell monolayer modelled by the AVM, implemented using the SAMoS software \cite{samos}.}
\label{fig1}
\end{figure}

Active vertex models (AVM) are currently widely used to describe and infer data from epithelial tissues, representing them as two dimensional (2D) networks of polygonal cells. The left panel in figure \ref{fig1} shows an image of a drosophila wing,
in which the cell junctions have been visualized by fluorescent labeling. 
This picture clearly motivates such a modeling approach, 
as sketched on the right panel.

The properties and interactions of the cells constituting the monolayer are implemented in the AVM by defining an energy function that typically reads
\begin{equation}
\mathcal{F} = \sum_{\alpha = 1}^N\left[ \frac{\kappa_{\alpha}}{2}(A_{\alpha} - A_0^\alpha)^2  + \frac{\Gamma_{\alpha}}{2}P_{\alpha}^2 \right] + \sum_{<\mu\nu>} \Lambda_{\mu \nu} l_{\mu \nu}\,.
\label{eq1}
\end{equation}
Here each cell is labeled by $\alpha = 1, ..., N$, and each vertex pair that shares a junction is designated by $\mu, \nu$.
The first term implements an area elasticity, with $\kappa_{\alpha}$ the modulus of cell area $A_{\alpha}$ around $A_0^\alpha$, its reference value. 
The second term is the perimeter contribution, with $\Gamma_{\alpha}$ 
the resistance of the cell to changing its perimeter $P_{\alpha}$. 
Finally, $\Lambda_{\mu \nu} = \gamma_c - \frac{\omega}{2}$ is the line tension of the cell junctions of length $l_{\mu\nu}$ that results from the cortical tension $\gamma_c$ along the contacts between cells and the cell-cell adhesion energy $\omega$. The implementation of the vertex model dynamics then rearranges the positions of all vertices, trying to minimize the energy function for a given set of parameters.

While the moduli $\kappa_\alpha$ and $\Gamma_\alpha$ are positive, $\Lambda_{\mu\nu}<0$. When the cell $\alpha$ shares junctions only with other cells of the same type, $\sum_{\langle\mu,\nu\rangle}\Lambda_{\mu\nu} l_{\mu\nu}=\Lambda_{\mu\nu}\sum_{\langle\mu,\nu\rangle}l_{\mu\nu}= \Lambda_{\mu\nu} P_\alpha$, and this term can be put together with the perimeter term, thereby yielding $\frac{\Gamma_\alpha}{2} (P_\alpha-P_0^\alpha)^2$ plus an unimportant constant, provided the target perimeter is $P_0^\alpha=-\Lambda_{\mu\nu}/\Gamma_\alpha>0$. The \textit{shape index} \cite{bi15},
\begin{equation}
p_0^{\alpha} = \frac{P_0^\alpha}{\sqrt{A_0^\alpha}} = \frac{|\Lambda_{\alpha\beta}|}{\Gamma_{\alpha}\sqrt{A_0^\alpha}}, \label{eq2}
\end{equation}
characterizes the ratio of the cell perimeter to the square root of its area. A critical value of this quantity is $p^{0*}~=~3.812$, which separates fluid-like and solid-like behavior of the tissue \cite{bi15,cza18}: for $p^0 < p^{0*}$, the monolayer is solid-like, and for $p^0 > p^{0*}$, it displays fluid-like behavior. Solid-like cells tend to be close to regular polygons and rarely give rise to fingering instabilities, whereas for fluid-like cells one finds both fingering instabilities and irregular cell shapes. 

In the standard implementation of the AVM,
the cells in the monolayer satisfy the following overdamped equations of motion \cite{bar17}
\begin{equation}
\zeta \dot{\boldsymbol{r}}_{\alpha} = f_a\boldsymbol{n}_{\alpha} + \boldsymbol{F}_{\alpha} + \nu_{\alpha},\label{eq3}
\end{equation}
and
\begin{equation}\label{eq4}
\zeta^r\dot{\theta}_{\alpha} = \boldsymbol{\tau}_{\alpha}\cdot \boldsymbol{N}_{\alpha}+\nu_{\alpha}^r.
\end{equation}
The unknowns $\boldsymbol{r}_{\alpha}$ and $\theta_{\alpha}$ are the positions of the centers of mass and the orientations of the directors of each cell $\alpha$, defined as $\boldsymbol{n}_{\alpha}=(\cos\theta_{\alpha},\sin\theta_{\alpha})$. In the 
center of mass equation,
$f_a\boldsymbol{n}_{\alpha}$
are active self-propulsion forces
along the vector $\bs{n}_{\alpha}$ determining the direction of cell motion,
$\bs{F}_{\alpha}$ are gradient
forces arising from the free energy function, Eq.~(\ref{eq1}), and  $\nu_{\alpha}$
stochastic forces, $\zeta$ is a friction coefficient.
In the angular equation,
$\boldsymbol{\tau}_{\alpha}$ and $\boldsymbol{N}_{\alpha}$
are the torque, stemming from cell-cell alignment models and
acting on 
the vector $\bs{n}_{\alpha}$,
and the normal vector to the cell monolayer (unit vector along the $z$-axis). $\nu_{\alpha}^r$
is a rotational noise and 
$\zeta^r$ the rotational friction.
Both noise terms are usually implemented as Gaussian white noise. 

In the following, we use a coarse-graining procedure that is Hamiltonian in nature. Hence we do not use Eqs.~(\ref{eq3}),(\ref{eq4}), but rather study a fluid of deformable particles, without active contributions. Nevertheless, the Poisson bracket approach accounts for the  dissipative contributions in the hydrodynamic limit and we will treat the vertex energy functional, Eq.~(\ref{eq1}), as faithfully as possible to keep track of the cellular nature of the system, especially the sensitivity to the shape index/line tension.   

\section{Coarse-graining procedure using Poisson brackets}\label{sec:3}

Our objective is to derive hydrodynamic equations for a fluid formed by deformable polygonal-shaped particles that is governed by the free energy of the AVM, 
Eq.~\eqref{eq1}. 
Unlike the AVM, we shall not include active forces in our equations. Thus, the equations will only reflect the effects of the fluid flow on cell shape and vice versa. 

\begin{figure}[t!]
	\centering
    \includegraphics[width=.8\linewidth]{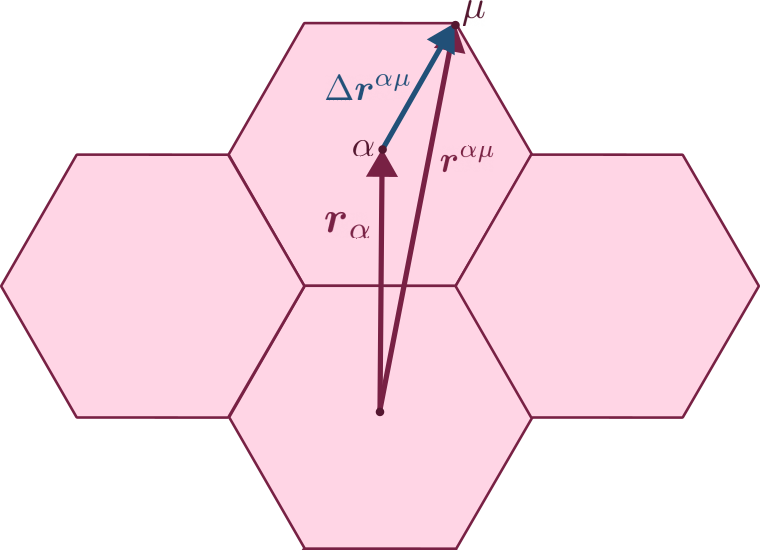}
    \caption{Definition of the vectors that are used to build up the shape tensor. The position of the cell center $\alpha$ is given by $\bs{r}_{\alpha}$, and the one of vertex $\mu$ of cell $\alpha$ is $\bs{r}^{\alpha \mu}$. Then, $\Delta  \bs{r}^{\alpha \mu} = \bs{r}^{\alpha \mu} - \bs{r}_{\alpha}$.} 
    \label{fig2}
\end{figure}

On large scales, this fluid is described by the average (coarse-grained) hydrodynamic fields $\phi^a$, which are the mass density $\rho$, the momentum density $\bs{g}$, and the cell-shape density tensor $G$. The latter accounts for the shape and elongation of the cells. The ``microscopic'' versions of the hydrodynamic fields are  \cite{her21}
\begin{eqnarray}
&&\hat{\rho} (\bs{r}, t) = \sum_{\alpha \mu} m^\alpha\delta(\bs{r} - \bs{r}^{\alpha \mu}(t)),\label{eq5}\\
&&\hat{\bs{g}}(\bs{r}, t) = \sum_{\alpha \mu} \bs{g}_{\alpha}\delta(\bs{r} - \bs{r}^{\alpha \mu}(t)),\label{eq6}\\
&&\hat{G}_{ij}^\alpha(\bs{r}, t) = \sum_{\alpha} G_{ij}^\alpha\delta(\bs{r} - \bs{r}^{\alpha}(t)),\label{eq7}\\
&&G_{ij}^\alpha=\frac{1}{n}\sum_{\mu=1}^n \Delta x_i^{\alpha\mu}\Delta x_j^{\alpha\mu}.\label{eq8}
\end{eqnarray}
Here $m^\alpha$ is the mass of cell $\alpha$ (we use $m^\alpha=1$)
and $\bs{r}^{\alpha \mu}$ 
the position of vertex $\mu$.
The momentum of cell $\alpha$ is given by
$\mathbf{g}_\alpha=m^\alpha\dot{\mathbf{r}}_\alpha$ 
and the shape tensor as specified via the vertices, with 
$\Delta x_i^{\alpha\mu}~=~~ \mathbf{r}^{\alpha\mu}~-~\mathbf{r}_\alpha$, where $\mathbf{r}_\alpha=\sum_\mu\mathbf{r}^{\alpha\mu}/n$ is the center of mass of the cell and Latin indices denote components, see Fig.~\ref{fig2}.
Note that the mass and momentum density fields are defined through the positions of the vertices in the delta function, 
while the shape tensor uses delta functions centered at the cell centers. This is due to the definition of the shape tensor in Eq. \eqref{eq8}, that already includes all vertices from each cell and is defined only for each cell center.

To obtain continuum equations for the hydrodynamic fields $\phi^\alpha$ from the Vertex model, we could resort to the Mori-Zwanzig projection technique \cite{zwa00}. However, although thought for different microscopic dynamics, it is easier to use the Poisson bracket approach, which is known to produce the usual hydrodynamics for non-deformable (and possibly anisotropic) particles \cite{cha95,sta03,maz06}. Given microscopic Hamiltonian dynamics, the evolution of a microscopic function $\hat{\phi}(\mathbf{r},t)$ obeys the equation
\begin{eqnarray}
    \frac{\partial \hat{\phi}}{\partial t} = \{ \hat{\mathcal{H}}, \hat{\phi}\} = \sum_{\alpha}\frac{\partial\hat{\mathcal{H}}}{\partial g_j^\alpha} \frac{\partial\hat{\phi}}{\partial r_j^\alpha}-\frac{\partial\hat{\phi}}{\partial g_j^\alpha}\frac{\partial\hat{\mathcal{H}}}{\partial r_j^\alpha},\label{eq9}
\end{eqnarray}
where the Hamiltonian function $\hat{\mathcal{H}}$ depends on the cellular positions and momenta $\mathbf{r}^\alpha$ and $\mathbf{g}^\alpha$. The coarse-graining operation consists of a spatial average on a length scale comprising many cells that is still small compared to macroscopic lengths. Given a coarse-grained free energy, $\mathcal{F}$, which depends on the macroscopic fields $\phi^a=\rho,\mathbf{g}, G$, the coarse-grained equations are \cite{cha95,her21}
\begin{eqnarray}
&&\frac{\partial \phi^a}{\partial t} = \{\mathcal{F}, \phi^a\} - \Gamma^{ab}\frac{\delta\mathcal{F}}{\delta \phi^b},\label{eq10}\\
&&\{\mathcal{F}, \phi^a\} = - \int d^2x' \mathcal{P}_{a b} (\bs{r}, \bs{r}') \frac{\delta \mathcal{F}}{\delta\phi^b(\bs{r}')}, \label{eq11}\\
&&\mathcal{P}_{a b}(\bs{r},\bs{r}') = \{ \phi^a (\bs{r}), \phi^b(\bs{r}')\} = - \mathcal{P}_{ba}(\bs{r}', \bs{r}),\label{eq12}
\end{eqnarray}
where summing over repeated indices is intended and we have dropped the time variable for now. The term $\{\mathcal{F}, \phi^a\}$ in Eq.~\eqref{eq10} yields the reactive part of the evolution equations and the other term yields the dissipative part. The latter is proportional to the generalized forces, as defined by variations of the Free energy, times kinetic coefficients $\Gamma^{ab}$ that obey  Onsager reciprocity relations near equilibrium \cite{deg84}. The coarse-grained Poisson brackets $\{\phi^a, \phi^b\}$, especially for the shape tensor, have been calculated previously in Ref.~\cite{her21}. 

The free energy can be split in a kinetic and a potential part 
\begin{eqnarray}
&&\mathcal{F}(\phi^a) = \int d^2x (f_K+f_V), \label{eq13}\\   
&& f_K= \sum_{i=1}^2\frac{g_i^2}{2\rho}= \sum_{i=1}^2\frac{1}{2}\rho v_i^2\,\,,\quad f_V= f_\text{sc}+f_\text{int}.\label{eq14}
\end{eqnarray}
The kinetic free energy is the usual one in terms of the average velocity $\mathbf{v}=\mathbf{g}/\rho$. The potential free energy is split into an average containing only single cell quantities, $f_\text{sc}$, and an interaction with neighboring cells, which, in analogy to nematic liquid crystals, we postulate to be
\begin{equation}
    f_\text{int}= \frac{K_G}{2} \left(\frac{\partial G_{kl}}{\partial x_i}\right)^2. \label{eq15}
\end{equation}
For a homogeneous phase with constant hydrodynamic fields, $f_K=f_\text{int}=0$, and the free energy density is $f=f_\text{sc}$. In the next section, we will find an expression for $f_\text{sc}$ from Eq.~(\ref{eq1}) in terms of the trace and deviatoric parts of the shape tensor. 

With the splitting of Eq.~\eqref{eq13}, the equations of motion \eqref{eq10} become \cite{her21}
\begin{eqnarray}
&&\frac{\partial\rho}{\partial t} +\bs{\nabla}\cdot(\rho\bs{v})=0,\label{eq16}\\
&&\rho \frac{d}{dt}v_i = -\partial_ip+\partial_j(\sigma_{ij}^D + \sigma_{ij}^E + \sigma_{ij}^G),\label{eq17}\\
&&\frac{D}{Dt}G_{ij} = G_{ik}D_{kj}+D_{ik}G_{kj}-\Gamma_{ijkl}\frac{\delta\mathcal{F}}{\delta G_{kl}}.\label{eq18}
\end{eqnarray}
Here $\partial_i=\partial/\partial x_i$, and
\begin{eqnarray}
&&\partial_iv_j = D_{ij} + \omega_{ij} + \frac{1}{2}\delta_{ij}\bs{\nabla}\cdot \bs{v}, \label{eq19}\\
&&D_{ij} = \frac{1}{2}(\partial_i v_j + \partial_j v_i - \delta_{ij}\bs{\nabla}\cdot \bs{v}),\label{eq20}\\
&&\omega_{ij} = \frac{1}{2}(\partial_iv_j - \partial_jv_i),\label{eq21}\\
&&\frac{d}{dt}=\frac{\partial}{\partial t}+ \mathbf{v}\cdot\nabla, \quad \frac{D}{Dt} = \frac{d}{dt} - [\omega,\cdot],\label{eq22}
\end{eqnarray}
are the gradient of the average velocity, the deviatoric part of its symmetrization (the rate of strain tensor), the vorticity, the material derivative and the co-rotational derivative, respectively. In the latter, $[A,B]_{ij}=A_{ik}B_{kj}-B_{ik}A_{kj} $. The continuity equation \eqref{eq16} does not contain a dissipative part. The pressure, and the different stress tensor contributions entering Eq.~\eqref{eq17}, are \cite{her21}
\begin{eqnarray}
&&p=\rho\frac{\delta\mathcal{F}_V}{\delta\rho} - f,\label{eq23}\\
&&\sigma_{ij}^D= 2\eta D_{ij}+\eta_b\delta_{ij}\bs{\nabla}\cdot\mathbf{v},\label{eq24}\\
&&\sigma_{ij}^E=-\frac{\partial f_\text{int}}{\partial(\partial_jG_{kl})}\partial_i G_{kl}= - K_G \partial_iG_{kl}\partial_jG_{kl},\label{eq25}\\
&&\sigma_{ij}^G= 2 G_{jk}\frac{\delta\mathcal{F}_V}{\delta G_{ik}}- \delta_{ij}G_{kl}\frac{\delta\mathcal{F}_V}{\delta G_{kl}}. \label{eq26}
\end{eqnarray}
For the sake of simplicity, in Eq.~(\ref{eq24}) we assumed that the dissipative part of the stress tensor, $\sigma_{ij}^D$, is that of an isotropic fluid with shear and bulk viscosity coefficients $\eta$ and $\eta_b$, respectively. In uniaxial anisotropic situations on expects five instead of just these two viscosities \cite{deg93}. The contributions $\sigma_{ij}^E$ and $\sigma_{ij}^G$ are reactive and the former corresponds to Erick stresses in liquid crystals \cite{deg93}. The coefficient tensor $\Gamma_{ijkl}$ in Eq.~\eqref{eq18} will be derived later, after we have analyzed the free energy density of the homogeneous phases.

\section{Homogeneous Free energy density of the vertex model}\label{sec:4}

Let us assume that the cells are regular (or almost regular) $n$-sided polygons. We also assume that $\kappa$, $A_0$, $\Gamma$ and $\Lambda$ are the same for all the cells and use $\sum_{<\mu\nu>} l_{\mu \nu} = P$ (perimeter). Then the cellular area and perimeter are given in terms of the shape tensor $G$ by \cite{her21}
\begin{eqnarray}
&&A_\alpha = \mu \sqrt{\text{det}(G)}, \quad P_\alpha = \nu \sqrt{ \text{Tr}(G)},\label{eq27}\\
&&\mu = \frac{n}{2}\sin\left(\frac{2\pi}{n}\right), \quad\nu = n\sqrt{2} \sin\left(\frac{\pi}{n}\right),\label{eq28}
\end{eqnarray}
with $n=6$ for hexagons.
For such a homogeneous phase with a single type of cells, Eq.~\eqref{eq1} results in the following expression for the vertex energy density
\begin{eqnarray}
f=\frac{\kappa}{2}(\mu\sqrt{\text{det}(G)}-A_0)^2\!  + \frac{\Gamma\nu^2}{2} \text{Tr}(G) \!+\! \Lambda \nu\sqrt{\text{Tr}(G)},\,\,\label{eq29}
\end{eqnarray}
where the cell area has been absorbed in the positive constants $\kappa$, $\Gamma$, and in $\Lambda<0$.

\begin{figure}[!t]
	\centering
    \includegraphics[width= 0.7\linewidth]{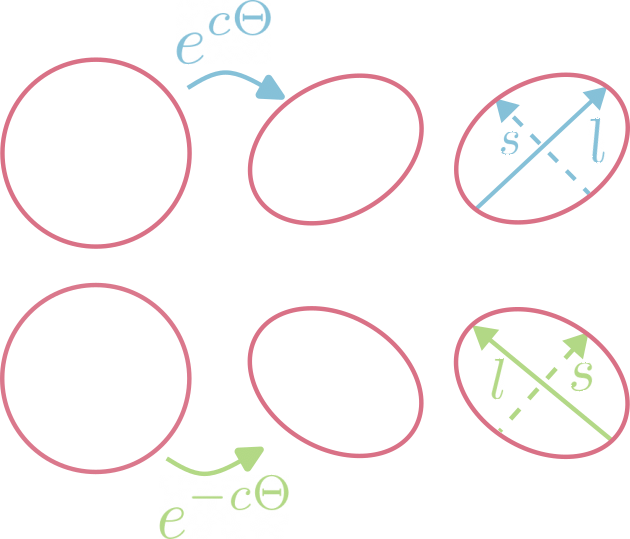}
    \caption{Sketch of the anisotropic states of a cell with $c~>~0$ (upper row) vs. $c~<~0$ (lower row). For a given orientation angle $\theta$, $c>0$ aligns the long axis of the ellipse, $l$, and the $\theta$ direction, whereas $c<0$ aligns the short axis of the ellipse, $s$, and the $\theta$ direction.} 
    \label{fig3}
\end{figure}

Following Ref.~\cite{ish17}, we write the shape tensor as
\begin{equation}
G = M_0 e^{c\Theta}= M_0 (\cosh{c}\, I+\sinh{c}\,\Theta), \label{eq30}
\end{equation}
where $M_0$ and $c$ are scalar fields, 
$I$ is the identity and 
$\Theta$ is the traceless symmetric tensor field 
\begin{equation}\label{eq31}
\Theta = 
\begin{pmatrix}
\cos{2\theta} & \sin{2\theta} \\
\sin{2\theta} & -\cos{2\theta}
\end{pmatrix}
.
\end{equation}
Note that $M_0^2=\text{det}(G)$ and 
$\Theta^2 = I$ hold. Importantly,
$c=0$ yields an isotropic shape tensor with  Tr$(G)=2\sqrt{\text{det}(G)}$. Thus, $c$ measures the anisotropy of the cells. The angle $\theta$ in Eq.~\eqref{eq31} determines the direction of the eigenvector corresponding to the largest eigenvalue of $G$. Fig.~\ref{fig3} sketches the anisotropic states of an elliptic cell with nonzero $c$. A related decomposition has been used directly on a triangular tiling of the plane representing a cellular tissue in Ref.~\cite{mer17}. However, Eq.~\eqref{eq30} is easier to relate to the free energy of Eq.~\eqref{eq29} written in terms of the average shape tensor.

In terms of $R=$ Tr$(G)$, we can rewrite Eq.~\eqref{eq30} as
\begin{equation}
G=\frac{R}{2}\, (I + \tanh c\Theta),\quad \tilde{G}= \frac{R}{2}\tanh{c}\Theta, \label{eq32}
\end{equation}
where the traceless tensor $\tilde{G}$ is the deviatoric part of the shape tensor. Thus, we can rewrite the energy density Eq.~\eqref{eq29} in the following simple form:
\begin{equation}
f = \frac{\kappa}{2}\left(\frac{\mu R}{2\cosh c}-A_0\right)^2\! + \frac{\Gamma\nu^2 R}{2} + \Lambda\nu \sqrt{R}\,. \label{eq33}
\end{equation}
Looking at Eq.~\eqref{eq32},
the fields $R$, $R\tanh c$ and $\theta$ occurring in the shape tensor parametrization describe the cell perimeter, the shape anisotropy and the director angle, respectively. As $R$ and $c$ appear naturally in the shape tensor and enter the free energy density, Eq.~\eqref{eq33}, there is no need to introduce a specific anisotropy field or to postulate a connection of the latter to the orientational order, 
as had been done previously in Ref.~\cite{her21}. 
Moreover, in Ref.~\cite{her21}
such a splitting was already done at the level of the shape tensor, calling for approximations already at the level of the Poisson brackets.

\begin{figure}[!b]
	\centering
    \includegraphics[width= \linewidth]{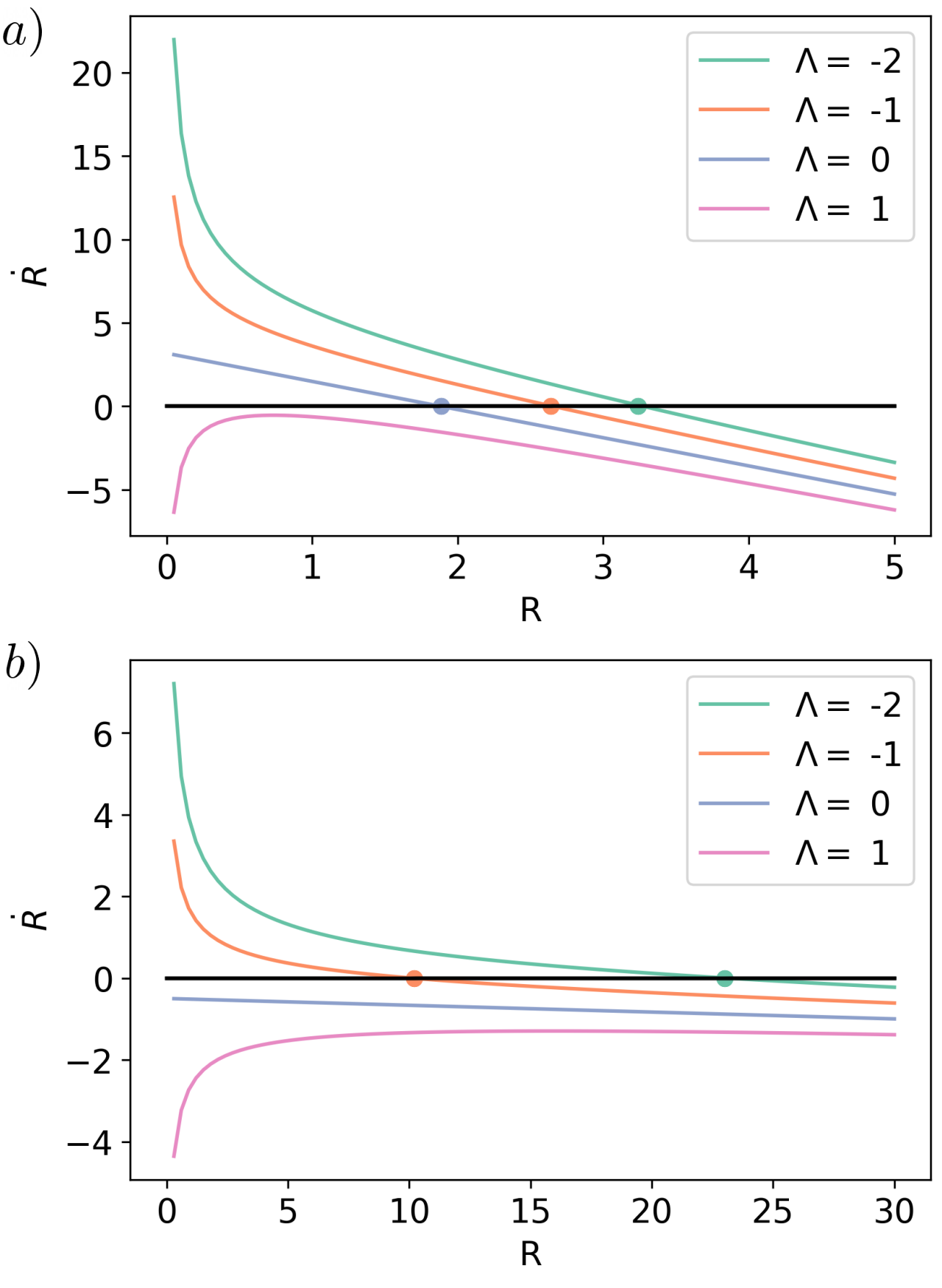}
    \caption{Graphical solution of $\partial f/\partial R=0$ by showing $\dot{R}=-\partial f/\partial R$ versus $R$ for different values of the line tension $\Lambda$. The parameters are $\kappa = 1$, $n = 6$ (hexagons), $\Gamma = 0.1$ and $A_0 = \pi$. a) Isotropic case, $c=0$, b) an anisotropic case with $c=3$.} 
    \label{fig4}
\end{figure}

\section{Thermodynamic stability}\label{sec:5}

The minima of the vertex Free energy, Eq.~\eqref{eq33}, correspond to stable homogeneous phases. At them, the first derivatives of $f$ vanish, which yields the conditions
\begin{subequations}\label{eq34}
\begin{eqnarray}
&&\frac{\partial f}{\partial R}\!=\!\frac{\kappa\mu}{2\cosh c}\!\left(\frac{\mu R}{2\cosh c}\!-\!A_0\right)\!+\frac{\nu}{2}\!\left(\Gamma\nu \!+\!\frac{\Lambda}{\sqrt{R}}\right)\!=\!0\,, \quad\quad\quad \label{eq34a}\\
&&\frac{\partial f}{\partial c}= -\frac{\kappa\mu R\sinh c}{2\cosh^2c}\!\left(\frac{\mu R}{2\cosh c}-A_0\right)\!=\!0\,. \label{eq34b}
\end{eqnarray}
\end{subequations}

\begin{figure}[b!]
	\centering
    \includegraphics[width= \linewidth]{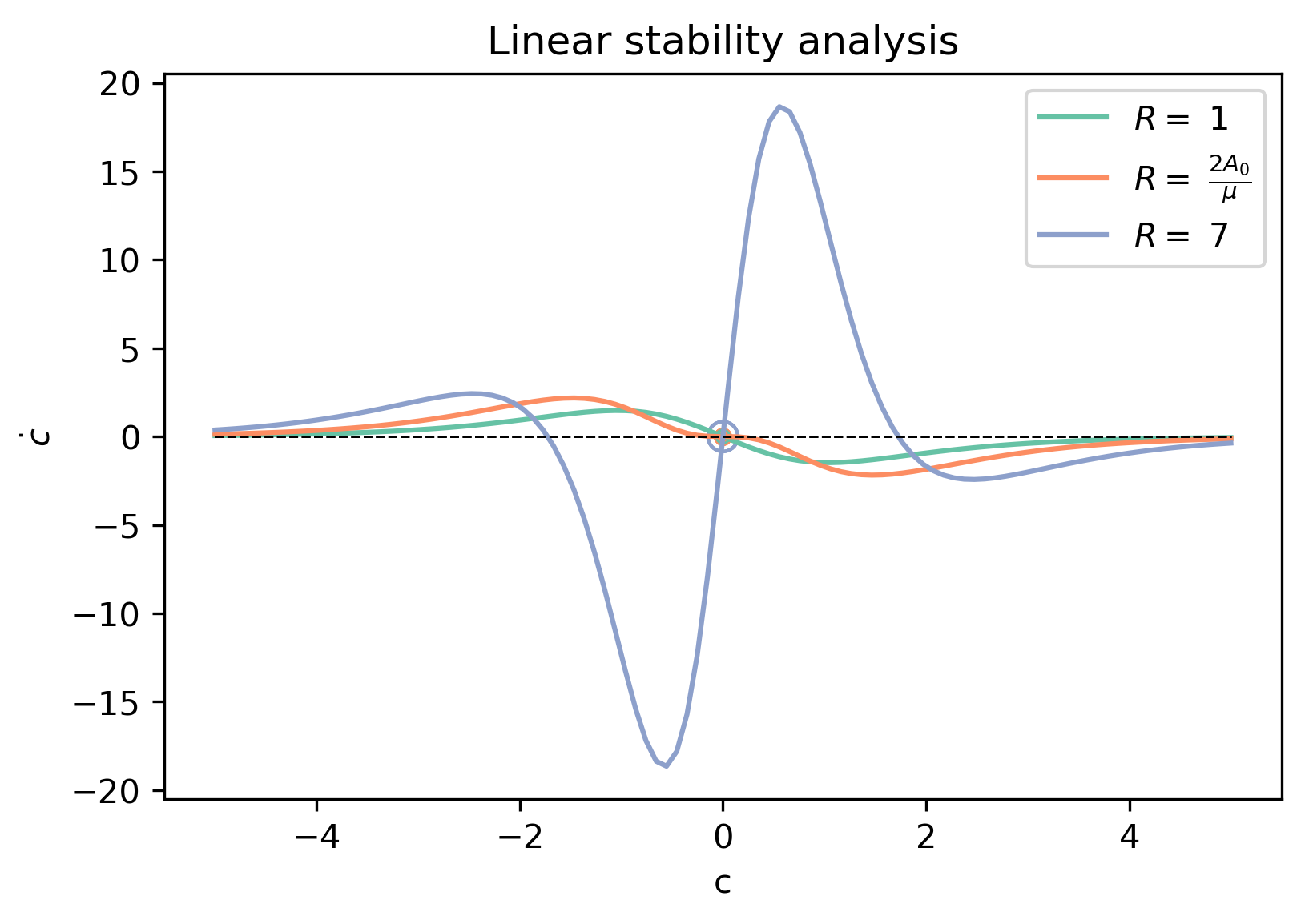}
    \caption{
    Graphical solution of $\partial f/\partial c=0$ by showing $\dot{c}=-\partial f/\partial c$ versus $c$ for different values of $R$,
    proportional to the squared perimeter.
    Shown are the three cases $R<R^*$, $R=R^*=2A_0/\mu$ and $R>R^*$.  The parameters are $\kappa = 1$, $n = 6$ (hexagons), $\Gamma = 0.1$ and $A_0 = \pi$.
    } 
    \label{fig5}
    \end{figure}

Fig.~\ref{fig4} shows graphically that solutions to Eq.~\eqref{eq34a} exist only if the line tension is non-positive  or, in the case of anisotropic phases, negative. In turn, Eq.~\eqref{eq34b} has isotropic solutions $c=0$ and $\pm c$-symmetric anisotropic solutions only for $R>2A_0/\mu$, as evident from the bracket term and also shown
graphically in Fig.~\ref{fig5}. Anisotropic solutions have $A=A_0$ and
hence from Eq.~\eqref{eq34a} it can be deduced that
$\sqrt{R}=|\Lambda|/(\Gamma\nu)$, i.e., $P=|\Lambda|/\Gamma$. These phases are stable if and only if the Hessian matrix of $f$ is positive definite, which is equivalent to
\begin{subequations}\label{eq35}
\begin{eqnarray}
&&\frac{\partial^2f}{\partial R^2}\!=\!\frac{\kappa\mu^2}{4\cosh^2c}-\frac{\nu\Lambda}{4R^\frac{3}{2}}>0\,,  \label{eq35a}\\
&&\frac{\partial^2f}{\partial c^2}= \frac{\kappa\mu R}{4\cosh^2c}[2A_0\cosh c(1-2\tanh^2c)\nonumber\\
&&\quad\quad\, - \mu R(1-3\tanh^2c)]>0\,, \label{eq35b}\\
&&\frac{\partial^2f}{\partial R^2}\frac{\partial^2f}{\partial c^2}-\!\left(\frac{\partial^2f}{\partial R\partial c}\right)^2\! >0\,. \label{eq35c}
\end{eqnarray}
\end{subequations}

\subsection{Cell shape instability and bifurcation}

Let us assume that the cell perimeter is constant and we want to ascertain whether phases with isotropic cells ($c=0$) are stable. If this is the case, $c=0$ has to be a minimum of the free energy. Writing $\sech c=\sqrt{1-\tanh^2c}\simeq 1-\frac{1}{2}\tanh^2c$ in Eq.~\eqref{eq33}, we obtain
\begin{eqnarray}
&&f-f_0\! \simeq \frac{\kappa\mu RA_0}{4}\!\left[\left(1\!-\!\frac{\mu R}{2A_0}\right)\!\tanh^2c+\frac{1}{4}\tanh^4c \right]\quad\quad\nonumber\\
&& \simeq\frac{\kappa\mu RA_0c^2}{4}\!\left[1-\frac{\mu R}{2A_0}+\frac{1}{3}\!\left(\frac{\mu R}{A_0}-\frac{5}{4}\right)c^2\right]\! \label{eq36}
\end{eqnarray}
up to order $\mathcal{O}(c^6)$ and
where $f_0$ is constant. Using Eq.~\eqref{eq27}, for fixed perimeter we get $R=P_0^2/\nu^2=\Lambda^2/(\Gamma^2\nu^4)$, and we obtain
\begin{eqnarray}
&&f=f_0+\frac{\kappa\mu^2P_0^2A_0c^2}{8\nu^4}\!\left[\frac{2\nu^2}{\mu}-\frac{P_0^2}{A_0} \right.\nonumber\\
&&\left. \quad
+\, \frac{1}{3}\!\left(\frac{2 P_0^2}{A_0}-\frac{5\nu^2}{2\mu}\right)c^2\right]\! +\, \mathcal{O}(c^6), \label{eq37}
\end{eqnarray}

\begin{figure}[b!]
	\centering
    \includegraphics[width= \linewidth]{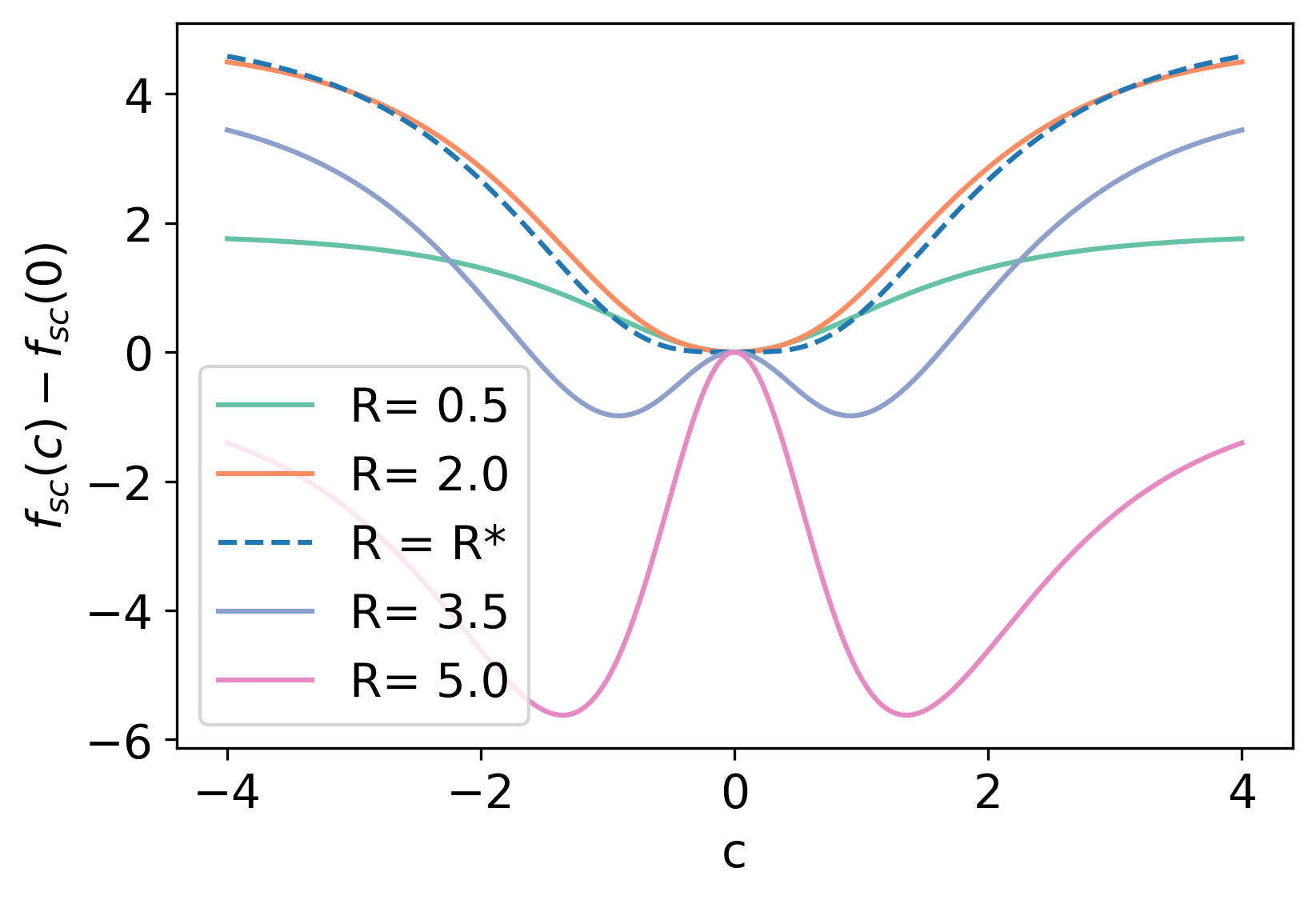}
    \caption{Normalized homogeneous free energy density versus anisotropy $c$ for different values of $R$. The dashed line corresponds to the critical value $R^*= 2A_0/\mu \approx 2.6$, as given by Eq.~(\ref{eq38}). For $R>R^*$, the isotropic state becomes unstable towards anisotropic cell shapes. Parameter values are $\kappa = 1$, $n = 6$, $\Gamma = 1.5$, $\Lambda = -1$, $A_0 = \pi$.} 
    \label{fig6}
\end{figure}

Clearly, the isotropic solution $c=0$ is stable when the shape index $p_0$ as defined in Eq.~\eqref{eq2} is smaller than a critical value given by
\begin{equation}
p_0^* = \frac{P_0}{\sqrt{A_0}} = \sqrt{\frac{2\nu^2}{\mu}}=\sqrt{4n\sin\!\left(\frac{\pi}{n}\right)\!}\,, \label{eq38}
\end{equation}
where we used Eq.~(\ref{eq28}).
At $p_0=p_0^*$, the free energy deviations around the isotropic phase are like $f-f_0\simeq\kappa\nu^2 A_0^2c^4/8\geq 0$. For $p_0<p_0^*$, the isotropic phase is stable, corresponding to a solid-like structure with cells being almost regular polygons. For $p_0>p_0^*$, a finite $c\neq 0$ emerges, corresponding to cells having irregular anisotropic shapes and the resulting configuration is fluid-like. Fig.~\ref{fig6} shows the free energy density, Eq.~(\ref{eq33}), as a function of $c$ for different values of $R$ and illustrates how the anisotropic phase appears.

We note that numerical simulations of the AVM 
\cite{her21} have given
the value $p_0^*\approx 3.81$. Interestingly, this corresponds to $n=5$, although pentagons cannot tile the plane. 
If we consider $n = 6$,  corresponding to a hexagonal tiling, we obtain $p_0^*\approx 3.72$, which slightly differs from the numerical simulations \cite{bi15}. 

\begin{figure}[!b]
	\centering
    \includegraphics[width= \linewidth]{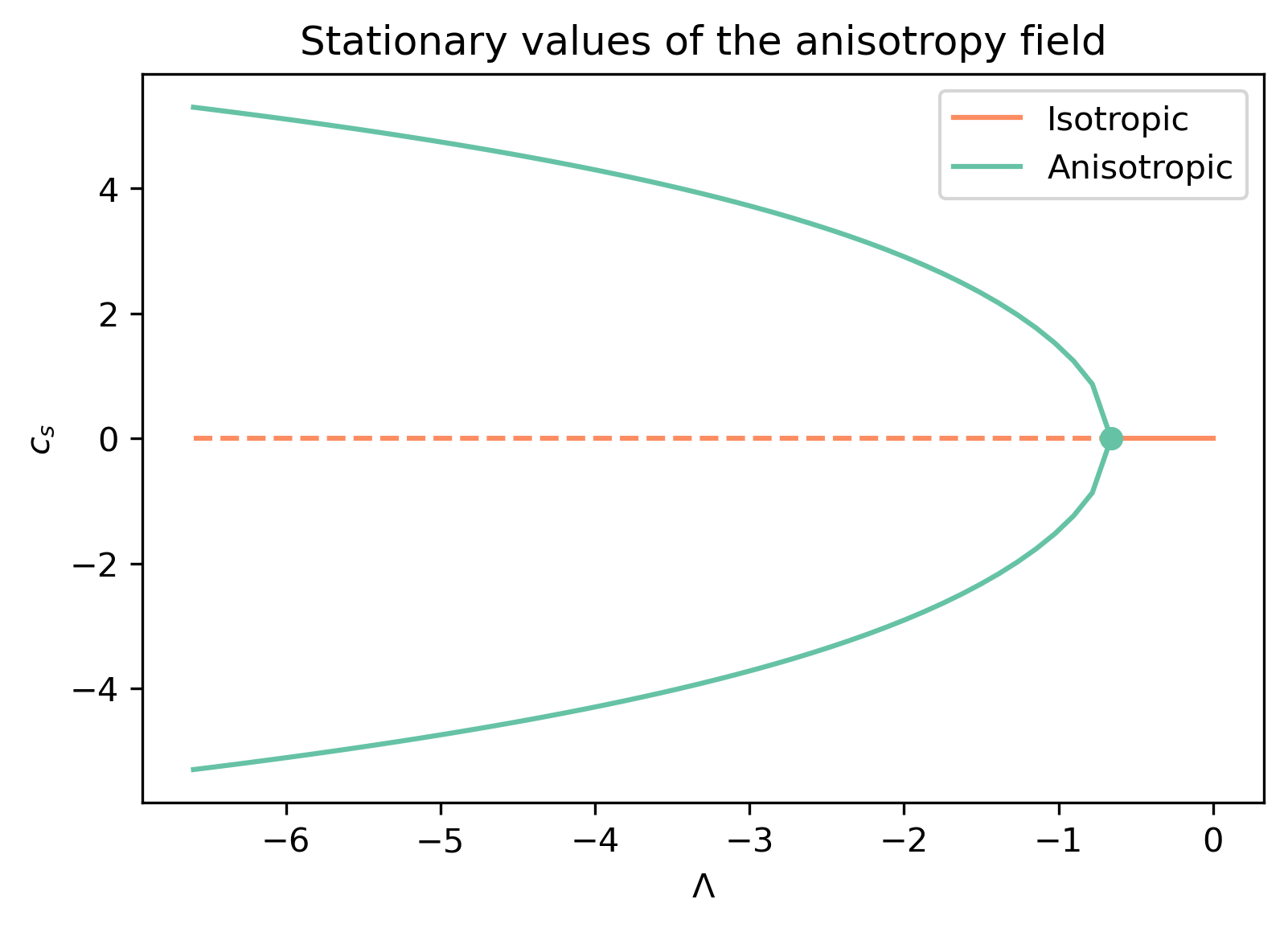}
    \caption{Pitchfork bifurcation  in the plane $(\Lambda,c)$,
    occurring at a critical line tension value $\Lambda^*~=~-\Gamma\nu\sqrt{2A_0/\mu}$.
    For $\Lambda>\Lambda^*$,
    the isotropic solution, $c=0$, is the only stationary state and stable.
    For $\Lambda<\Lambda^*$, the isotropic state becomes unstable (dashed line) and the
    system chooses one of the symmetric anisotropic branches.} 
    \label{fig7}
\end{figure}

Eq.~\eqref{eq37} is an approximation of Eq.~\eqref{eq33} which, for fixed perimeter $P_0=|\Lambda|/\Gamma$ and $\Lambda<0$ becomes
\begin{equation}
f= \frac{\kappa\mu^2\Lambda^4}{8\Gamma^4\nu^4}\left(\sech c - \frac{2A_0\Gamma^2\nu^2}{\mu\Lambda^2}\right)^2\!-\frac{\Lambda^2}{2\Gamma}.\label{eq39}
\end{equation}
Similarly, fixing the area to $A_0$, $\sqrt{R}=\sqrt{2A_0\cosh c/\mu}$, and Eq.~\eqref{eq33} becomes
\begin{equation}
f= \frac{\Gamma\nu^2A_0}{\mu}\left(\sqrt{\cosh c} - \sqrt{\frac{\mu\Lambda^2}{2A_0\Gamma^2\nu^2}}\right)^2\!-\frac{\Lambda^2}{2\Gamma}.\label{eq40}
\end{equation}

For $p^0>p^{0*}$, the values of $c$ that solve Eq.~\eqref{eq34b} are either $c=0$, or 
\begin{equation}
\cosh c=
\frac{\mu\Lambda^2}{2A_0\Gamma^2\nu^2}. \label{eq41}
\end{equation}

Solutions of this equation with $c\neq 0$ correspond to anisotropic phases, which are stable, as evident according to either Eqs.~\eqref{eq39}, \eqref{eq40}, or the criteria \eqref{eq35}, which simply become: 
\begin{eqnarray*}
&&\frac{\partial^2f}{\partial R^2}=\frac{\kappa A_0^2}{R^2}+\frac{\Gamma\nu^2}{4R}  >0\,,\,\,
\frac{\partial^2f}{\partial c^2}= \kappa A_0^2\tanh^2c >0, \\
&&\frac{\partial^2f}{\partial R^2}\frac{\partial^2f}{\partial c^2}-\!\left(\frac{\partial^2f}{\partial R\partial c}\right)^2\!= \frac{\kappa\Gamma\nu^2A_0^2}{4R} \tanh^2c >0. 
\end{eqnarray*}

We can now draw
the bifurcation diagram of anisotropy $c$ versus the 
line tension $\Lambda$ as the control parameter, with its
critical value $\Lambda^*=-\Gamma\nu\sqrt{2A_0/\mu}$
obtained from Eq.~(\ref{eq38}). The bifurcation diagram, displaying a pitchfork bifurcation, is shown in Fig.~\ref{fig7}. Cells with $c~>~0$ and $c~<~0$ have the same energy (since there is no preferred direction in the free energy), but different orientations, cf.~Fig \ref{fig3} 
\cite{ish17, mer17}. 

\subsection{Cell area instability of the isotropic phase}
It is also interesting to find the inflection point of the free energy density,  Eq.~\eqref{eq33}, as a function of $R$ for the isotropic phase. 
For $c=0$, from Eq.~\eqref{eq35a}, one obtains the stability condition
\begin{subequations}\label{eq42}
    \begin{eqnarray}
\frac{\kappa \mu^2}{4}-\frac{\nu\Lambda}{4R^\frac{3}{2}}>0\Longrightarrow\frac{\Lambda}{\kappa A^\frac{3}{2}}<\sqrt{\frac{\mu}{2\nu^2}}=\frac{1}{p_0^{*}}. \label{eq42a}
\end{eqnarray}
From Eqs.~\eqref{eq27}, \eqref{eq30} with $c=0$, $A=\mu M_0=\mu R/2$. Then Eq.~\eqref{eq35b} with $c=0$ yields
\begin{equation}A < A_0 .\label{eq42b}
\end{equation}
\end{subequations}

\begin{figure}[!t]
	\centering
    \includegraphics[width= \linewidth]{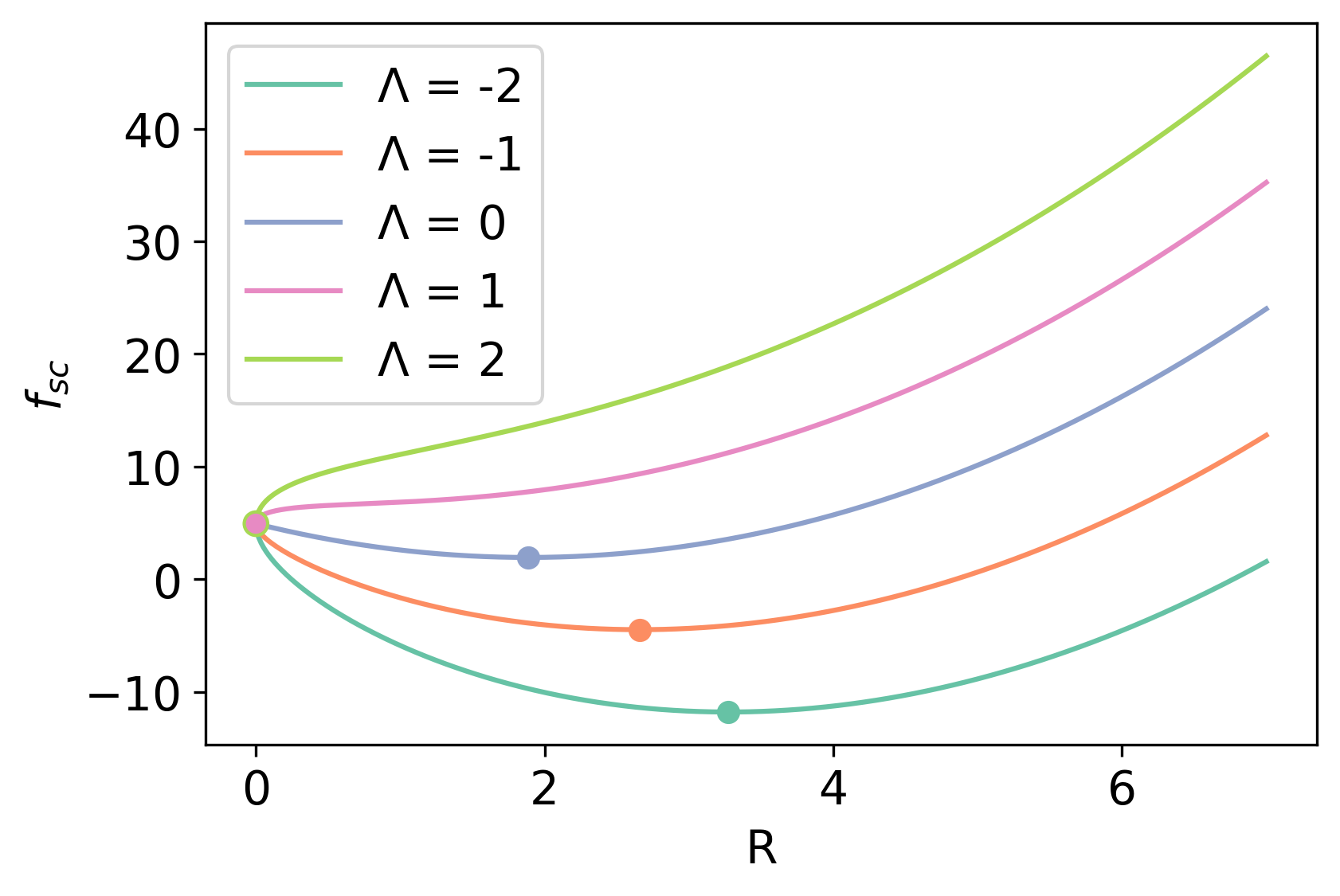}
    \caption{Free energy density of the isotropic homogeneous state ($c=0$) as a function of $R$ for different line tension values $\Lambda$, with $\kappa = 1$, $n = 6$ (hexagons), $\Gamma = 0.1$, $A_0 = \pi$. The minimum of the free energy at $R^*>0$ disappears for sufficiently large positive values of the line tension.} 
    \label{fig8}
\end{figure}

Hence there are two ways in which the isotropic phase may become unstable: 
For sufficiently negative line tension,
if the area reaches the target area $A_0$, anisotropic phases emerge from the isotropic phase as shown in Figs.~\ref{fig6} and \ref{fig7}. In turn, if the scaled line tension surpasses  $1/p_0^{*}$, the homogeneous isotropic phase becomes unstable,
 cf.~Fig.~\ref{fig8},
but homogeneous anisotropic phases are not stable either.
While Fig.~\ref{fig8} seems to suggest that the cells shrink to zero area, it could also happen that spatially non-homogeneous phases may appear.\\

\section{Kinetic coefficients and equation of motion for the shape tensor}\label{sec:6}

Now that we know the behavior
of the vertex Free energy in terms of the variables $R$ and $c$ introduced in the parametrization of the shape tensor,
let us come back to the coarse graining procedure
of section \ref{sec:3}.
As discussed above, 
since in general $G_{ij}$ varies in space, we have to add the term $f_\text{int}$ of Eq.~\eqref{eq15} to the free energy density that penalizes gradients, where 
for simplicity we used a one-constant-approximation 
(cf.~Frank elasticity in nematic liquid crystals \cite{LL7,deg93}). 
Then, the coarse-grained free energy density, Eq.~(\ref{eq33}) can be written as
\begin{eqnarray}
&&f_V\!= \frac{\kappa}{2}\!\left(\frac{\mu R}{2\cosh c}\!-\!A_0\right)^2\!   + \frac{\Gamma\nu^2\! R}{2}+\Lambda \nu\sqrt{R} +\! \frac{K_G}{4}|\bs{\nabla}R|^2\! \nonumber\\
&&\quad + \frac{K_G}{4} \left[|\bs{\nabla}(R\tanh c)|^2+4(R\tanh c|\bs{\nabla}\theta|)^2\right].  \label{eq43}
\end{eqnarray}
\begin{widetext}
To decide the form of the kinetic coefficients $\Gamma_{ijkl}$ in Eq.~\ref{eq18}, we first derive the following formulas
\begin{subequations}\label{eq44}
\begin{eqnarray}
&&\frac{D}{Dt}G_{ij}= \frac{\delta_{ij}}{2}\frac{dR}{dt}+ \frac{\Theta_{ij}}{2}\frac{d}{dt}(R\tanh c)+\frac{R}{2}\tanh c \left(\frac{d\theta}{dt}+
\frac{\partial_1 v_2- \partial_2 v_1}{2}\right)\frac{\partial\Theta_{ij}}{\partial\theta},\label{eq44a}\\
&& D_{ik}G_{kj}+G_{ik}D_{kj} = \frac{R}{2}\tanh\!c\left[(\partial_1v_1-\partial_2v_2)\cos 2\theta+(\partial_1v_2+\partial_2v_1)\sin 2\theta\right] \delta_{ij} +RD_{ij}, \label{eq44b}\\
&&\frac{D}{Dt}G_{ij}-D_{ik}G_{kj}-G_{ik}D_{kj} = \frac{\delta_{ij}+\Theta_{ij}\tanh c}{2}\left(\frac{dR}{dt}-\big[(\partial_1v_1-\partial_2v_2)\cos 2\theta+(\partial_1v_2+\partial_2v_1)\sin 2\theta\big]R\tanh\!c\right)\! \nonumber\\
&&\quad\quad\quad + \frac{R\Theta_{ij}}{2}\left(\frac{1}{\cosh^2c}\frac{dc}{dt} +\big[(\partial_1v_1-\partial_2v_2)\cos 2\theta+(\partial_1v_2+\partial_2v_1)\sin 2\theta\big]\tanh^2\!c \right)\!\nonumber\\
&&\quad\quad\quad+\frac{R}{2}\tanh c \left(\frac{d\theta}{dt}+\frac{\partial_1 v_2- \partial_2 v_1}{2}\right)\frac{\partial\Theta_{ij}}{\partial\theta} - RD_{ij} = - \Gamma_{iikl} \frac{\delta\mathcal{F}}{\delta G_{kl}}
,\label{eq44c}\\
&&\frac{\partial}{\partial G_{kl}}\text{Tr}G = \delta_{kl},\quad
\frac{\partial}{\partial G_{kl}}\text{det}G= (\text{det}G) G_{kl}^{-1} =R\delta_{kl}-G_{kl}=\frac{R}{2}\delta_{kl}-\frac{R\tanh c}{2}\Theta_{kl},\label{eq44d}
\end{eqnarray}\end{subequations}
where \eqref{eq44c} is just an expanded version of \eqref{eq18}. 
Using Eq.~\eqref{eq44d}, from Eq.~\eqref{eq43} we  obtain 
\begin{eqnarray}
\frac{\delta\mathcal{F}}{\delta G_{kl}}\!&=&\! \frac{\delta_{kl}}{2}\!\left[\kappa\mu\!\left(\frac{\mu R}{2}-A_0\cosh c\right)\! + \Gamma\nu^2+\frac{\Lambda\nu}{\sqrt{R}}-K_G\nabla^2R\right]\!-\Theta_{kl}\! \left[\frac{\kappa\mu}{2}\!\left(\frac{\mu R}{2\cosh c}-A_0\right) \sinh c\right. \nonumber\\ 
&+&\!\left. \frac{K_G}{2}(\nabla^2(R\tanh c)-4R\tanh c\,|\bs{\nabla}\theta|^2) \right]\! - \frac{K_G}{2} [R\tanh c\, \nabla^2\theta+2\bs{\nabla}\theta\cdot\bs{\nabla}(R\tanh c)]\frac{\partial\Theta_{kl}}{\partial\theta}.  \label{eq45}
\end{eqnarray}
\end{widetext}

Note that the the matrices $\delta_{ij}$, $\Theta_{ij}$ and $\partial\Theta_{ij}/\partial\theta$ appearing in Eqs.~\eqref{eq44} and \eqref{eq45} are orthogonal with respect to the scalar product Tr$(A_{ik}B_{kj})$ and form a basis in the corresponding vector space. Thus, the kinetic coefficients can be written as linear combinations of products of these matrices.  We will select them by imposing that, at zero average flow velocity, $v=0$, Eq.~\eqref{eq18} should yield a gradient system for homogeneous phases compatible with the thermodynamic stability established in the last section. 

According to Eq.~\eqref{eq44c}, the evolution equation for $R$ can be found by taking the trace, thereby obtaining
\begin{eqnarray}
\frac{dR}{dt}\!&=&\! R\tanh{c}\big[\left(\partial_1v_1-\partial_2v_2\right)\cos{2\theta} \nonumber \\
&+&\! \left(\partial_1v_2  + \partial_2v_1\right)\sin{2\theta}\big]
- \Gamma_{iikl} \frac{\delta\mathcal{F}}{\delta G_{kl}}, \label{eq46}
\end{eqnarray}
For Eq~\eqref{eq46} to be a gradient vector field for zero velocity, we consider
\begin{equation*}
\frac{\delta \mathcal{F}}{\delta R} = \frac{\delta \mathcal{F}}{\delta G_{kl}}\frac{\partial G_{kl}}{\partial R}
= \frac{\delta \mathcal{F}}{\delta G_{kl}}\frac{G_{kl}}{R}.
\end{equation*}
Thus, we should select $\Gamma_{iikl}=\gamma_1 G_{kl}/R=\gamma_1 R(\delta_{kl}+\tanh c\,\Theta_{kl})/2$.

Similarly, $\dot{c}$ in Eq,~\eqref{eq44a} has a prefactor $R\Theta_{ij}/(2\cosh^2c)$ and, therefore, the kinetic coefficient in its equation has to produce the gradient vector field $-\gamma_2 R\Theta_{ij} [\delta\mathcal{F}/\delta c]/(2\cosh^2c)$, where
\begin{eqnarray*}
\frac{\delta \mathcal{F}}{\delta c}\!& =&\! \frac{\delta \mathcal{F}}{\delta G_{kl}}\frac{\partial G_{kl}}{\partial c}
= \frac{\delta \mathcal{F}}{\delta G_{kl}}\frac{R\, \Theta_{kl}}{2\cosh^2{c}}\\
\!&=&\!\frac{\partial f_\text{sc}}{\partial c}
 -\frac{K_GR}{2\cosh^2c}[\nabla^2(R\tanh c) - 4R\tanh c|\bs{\nabla}\theta|^2], 
\end{eqnarray*}
The first line suggests a second contribution to the kinetic coefficient. We choose a third one, as explained below, and write
\begin{eqnarray}
\Gamma_{ijkl}\!= \gamma_1\frac{G_{ij}G_{kl}}{R^2} + \gamma_2 \frac{R^2\Theta_{ij}\Theta_{kl}}{4\cosh^4{c}} + \frac{\gamma_3}{8} \frac{\partial\Theta_{ij}}{\partial\theta}\frac{\partial\Theta_{kl}}{\partial\theta}.\quad \label{eq47}
\end{eqnarray}
Note that the kinetic coefficients $\Gamma_{ijkl}$ have to be symmetric with respect to the exchanges $ij \leftrightarrow ji$ and in $kl \leftrightarrow lk$ (because stress and shear are symmetric tensors) and $ij \leftrightarrow kl$ (Onsager relation) \cite{deg84}, which is all fulfilled by our choice. 

Looking now at the traceless part of Eq.~\eqref{eq18} in the form of Eq.~\eqref{eq44c}, using
 \eqref{eq46} and \eqref{eq47}, we get
\begin{subequations}\label{eq48}
\begin{eqnarray}
    A\Theta_{ij}+ \frac{B}{2} \frac{\partial\Theta_{ij}}{\partial\theta}= R D_{ij},\label{eq48a}
\end{eqnarray}
where $A$ and $B$ are the coefficients of $\Theta_{ij}$ and of $(\partial\Theta_{ij}/\partial\theta)/2$ in Eqs.~\eqref{eq44c} and \eqref{eq18}:
\begin{eqnarray}
&&A=\frac{R}{2\cosh^2\!c}\!\left(\frac{dc}{dt} +\gamma_2\frac{\partial f_{sc}}{\partial c}
+\mathcal{A}\sinh^2\!c\right. \nonumber\\
&&\quad\left.- \gamma_2K_G[\nabla^2(R\tanh\!c)-4R\tanh c|\bs{\nabla}\theta|^2]\right)\!, \label{eq48b}\\
&&B= R\tanh\! c\left(\frac{d\theta}{dt}+\frac{\partial_1v_2-\partial_2v_1}{2}-\gamma_3K_G[\nabla^2\theta\right.\nonumber\\
&&\quad\left.+2\bs{\nabla}\theta\cdot\bs{\nabla}\ln(R\tanh c)] \right)\!,  \label{eq48c}\\
&&\mathcal{A}=(\partial_1v_1-\partial_2v_2)\cos 2\theta+ (\partial_1v_2+\partial_2v_1)\sin 2\theta.\quad\quad \label{eq48d}
\end{eqnarray}
\end{subequations}
 Eq.~\eqref{eq48a} is equivalent to 
\begin{widetext}
\begin{eqnarray}
\left(A\cos 2\theta-B\sin 2\theta-R\frac{\partial_1v_1-\partial_2v_2}{2}\right)
\begin{pmatrix}
1 & 0 \\
0 & \quad -1
\end{pmatrix} = -\left(A\sin 2\theta+B\cos 2\theta-R\frac{\partial_1v_2+\partial_2v_1}{2}\right)
\begin{pmatrix}
0 & 1 \\
1 & 0
\end{pmatrix}. \label{eq49}
\end{eqnarray}
The coefficients of the independent matrices in Eq.~\eqref{eq49} have to be zero, which allows
to get the equations for $\frac{dc}{dt}$ and $\frac{d\theta}{dt}$ from (\ref{eq48}).
Together with Eq.~\eqref{eq46}, the final equations are
\begin{subequations}\label{eq50}
\begin{eqnarray}
\frac{dR}{dt}\!&=&\!R\tanh{c}\,[\left(\partial_1v_1\!-\partial_2v_2\right)\cos{2\theta}+\! \left(\partial_1v_2  + \partial_2v_1\right)\sin{2\theta}]
- \gamma_1\frac{\partial f_\text{sc}}{\partial R}\nonumber\\
&+&\!\frac{\gamma_1K_G}{2}\Big[\nabla^2R 
 + \tanh{c}\nabla^2(R\tanh{c})-4R\tanh^2{c}|\bs{\nabla}\theta|^2\Big]\!, \label{eq50a}\\
 \frac{dc}{dt}\!&=&\!(\partial_1v_1 - \partial_2v_2)\cos 2\theta + (\partial_1v_2  + \partial_2v_1)\sin 2\theta - \gamma_2\frac{\partial f_\text{sc}}{\partial c} +\frac{\gamma_2K_GR}{2\cosh^2c}\Big[\nabla^2(R\tanh{c})-4R\tanh{c}|\bs{\nabla}\theta|^2\Big],\, \label{eq50b}\\
 \frac{d\theta}{dt}\!&=&\!-\frac{\partial_1v_2  - \partial_2v_1}{2}+\frac{(\partial_1v_2  + \partial_2v_1)\cos 2\theta - (\partial_1v_1 - \partial_2v_2)\sin 2\theta}{2\tanh c}+\gamma_3 K_G \,\left[\nabla^2\theta 
 + 2\bs{\nabla}[\ln(R\tanh{c})]\cdot\bs{\nabla}\theta\right]\!. \label{eq50c}
\end{eqnarray}\end{subequations}
\end{widetext}

In the equation for $\dot{\theta}$ we used that
\begin{eqnarray*}
    \frac{\delta \mathcal{F}}{\delta G_{kl}} \frac{\partial\Theta_{kl}}{\partial\theta} = - 4 K_G\left[
    R\tanh c\, \nabla^2\theta+ 2 \bs{\nabla}(R\tanh{c})\cdot\bs{\nabla}\theta\right]\!.
\end{eqnarray*}
Hence the choice of the contribution proportional to $\gamma_3$ in Eq.~\eqref{eq47} leads to 
Eq.~\eqref{eq50c} becoming a diffusion equation for the angle.

\section{Homogeneous phases at zero flow velocity}\label{sec:7}
For homogeneous phases at zero velocity, $R$ and $c$ depend only on time and Eqs.~\eqref{eq50} become
(putting all contributions in $v$ and $K_G$ to zero)
\begin{subequations}\label{eq51}
\begin{eqnarray}
&&\dot{R} =- \gamma_1 \frac{\partial f_\text{sc}}{\partial R}=- \frac{\gamma_1\kappa \mu}{2\cosh{c}} \left( \frac{\mu R}{2\cosh{c}} - A_0 \right) \nonumber\\
 &&\quad\, - \frac{\gamma_1 }{2} \left(\frac{\Lambda \nu}{\sqrt{R}} +\Gamma \nu^2\right)\!, \label{eq51a}\\
&&\dot{c} =- \gamma_2 \frac{\partial f_\text{sc}}{\partial c}=  \gamma_2 \frac{R\kappa \mu\tanh{c}}{2\cosh{c}}\left( \frac{\mu R}{2\cosh{c}} - A_0 \right)\!,\quad\quad \label{eq51b}\\
&&\dot{\theta}= 0. \label{eq51c}
\end{eqnarray}   
\end{subequations}

\begin{figure}[b!]
    \centering
    \includegraphics[width=.95\linewidth]{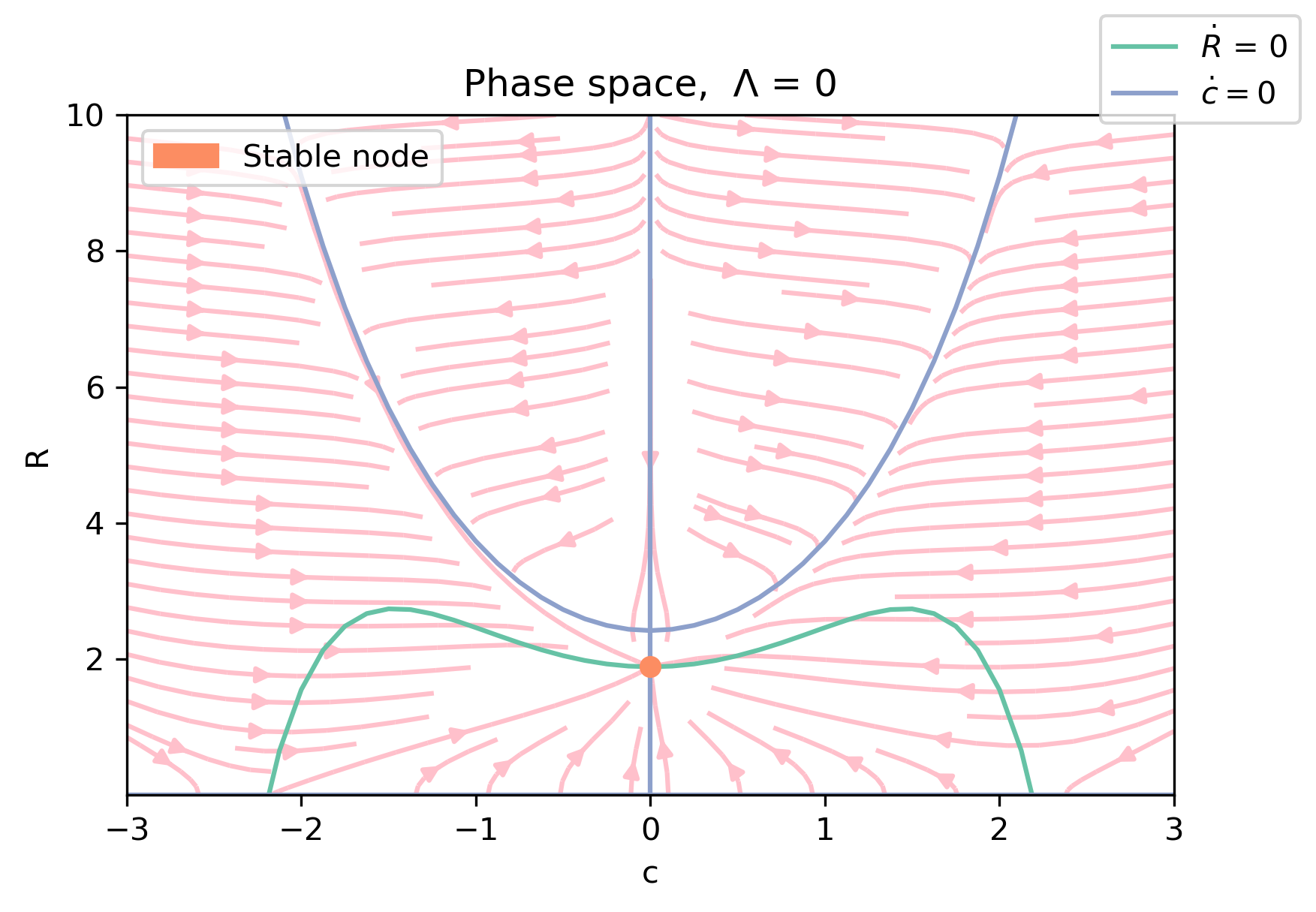}
    \caption{Phase space portrait of Eqs.~\eqref{eq51} for $\Lambda = 0$. The only stationary solution occurs at $c=0$ (isotropic phase) at finite $R$ (orange dot). Shown are the nullclines (blue and green curves) and the streamlines of the dynamical system (red arrows). 
    Here $\kappa = 1$, $n = 6$, $\Gamma = 0.1$, $A_0 = \pi$.} 
    \label{fig9}
\end{figure} 

\begin{figure}[!b]
	\centering
    \includegraphics[width=.95\linewidth]{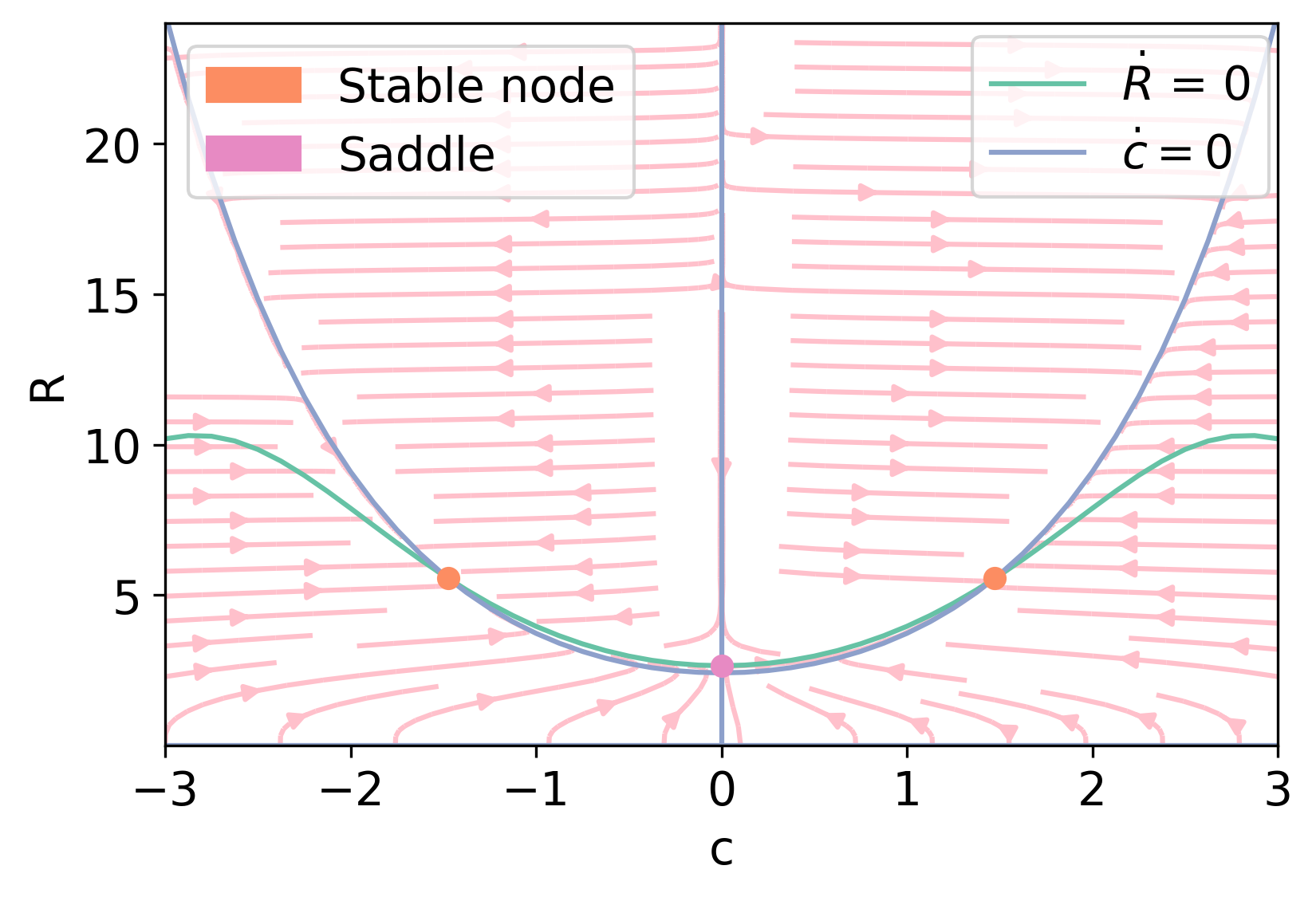}
    \caption{Same as Fig.~\ref{fig9} for $\Lambda=-1<\Lambda^*$. There are two $\pm c$-symmetric stable nodes (orange dots) corresponding to the 
    anisotropic phases. The isotropic phase at $c=0$ (purple dot) becomes a saddle point.  } 
    \label{fig10}
\end{figure}

The stationary solutions of Eqs.~\eqref{eq51} are 
-- by construction, i.e.~by the choice of the dissipative coefficients -- exactly
the homogeneous phases
already discussed in  Section \ref{sec:5}. 
Their linear stability depends on the eigenvalues of the Jacobian matrix
\begin{equation}\label{eq52}
J = 
\begin{pmatrix}
-\gamma_1\frac{ \partial^2f_\text{sc}}{\partial R^2} & -\gamma_1\frac{\partial^2f_\text{sc}}{\partial R\partial c}\\
 -\gamma_2\frac{\partial^2f_\text{sc}}{\partial R\partial c} & 
-\gamma_2\frac{\partial^2f_\text{sc}}{\partial c^2}
\end{pmatrix}.
\end{equation}

The stability criteria in Eqs.~\eqref{eq35} (corresponding to a positive definite Hessian matrix of the free energy density $f_\text{sc}$) ensure that the eigenvalues of the Jacobian \eqref{eq52} are negative and, therefore, that the homogeneous phases are dynamically stable. This is further illustrated by the phase space portraits \cite{str94} of the dynamical system defined by Eqs.~\eqref{eq51a} and \eqref{eq51b}. If $\Lambda>\Lambda^*=-\Gamma\nu\sqrt{2A_0/\mu}$, the only stationary solution is isotropic and a stable node as shown in Fig.~\ref{fig9}. For $\Lambda<\Lambda^*$, the isotropic solution becomes an unstable saddle point and there are two symmetric anisotropic solutions that are stable nodes, as illustrated in Fig.~\ref{fig10}. The anisotropic phases emerge from the homogeneous phase at $\Lambda=\Lambda^*$, cf.~the pitchfork bifurcation shown in Fig.~\ref{fig7}. 

The area $A=\mu R/(2\cosh c)$ calculated from homogeneous phases, i.e., stationary solutions  of Eqs.~\eqref{eq51}, increases with negative line tension, from $\Lambda=0$ to $-\Lambda=|\Lambda^*|$. Then $A=A_0$ in the anisotropic phase for all $\Lambda\leq \Lambda^*$.

\section{Homogeneous phases under shear flow}\label{sec:8}
Let us now consider the system under a stationary homogeneous shear flow
\begin{equation}
v_y = 0, \quad v_x = \dot{\gamma}y, \label{eq53}
\end{equation}
that solves the continuity equation, Eq.~\eqref{eq16} and, to leading order, the velocity equation, Eq \eqref{eq17}
(i.e.~including $\sigma^D$ and $\sigma^E$, but neglecting a possible feedback of the ordering on the flow as
described by $\sigma^G$).

Substituting Eq,~\eqref{eq53} into Eqs.~\eqref{eq50}, they become
\begin{subequations} \label{eq54}
\begin{eqnarray}
&&\dot{R} = \dot{\gamma} R \tanh{c} \sin{2\theta} - \gamma_1 \frac{\partial f_\text{sc}}{\partial R},\label{eq54a}\\
&&\dot{c} = \dot{\gamma}\sin{2\theta} - \gamma_2 \frac{\partial f_\text{sc}}{\partial c},\label{eq54b}\\
&& \dot{\theta} = \frac{\dot{\gamma}}{2}\left(1+\frac{\cos 2\theta}{\tanh c}\right)\!. \label{eq54c}
\end{eqnarray}
\end{subequations}
Note that uniform shear introduces a relation between the "director orientation" at angle $\theta$ and the cellular anisotropy and perimeter fields, $c$ and $R$, although the average free energy does not depend on $\theta$.

\subsection{Stationary solutions}
The stationary solutions of Eqs.\eqref{eq54} satisfy the following system of equations:
\begin{subequations} \label{eq55}
\begin{eqnarray}
&&\dot{\gamma} R\tanh{c}\sin{2\theta} =  \frac{\gamma_1}{2}\bigg[ \frac{\kappa \mu}{ \cosh{c}}\left(\frac{\mu R}{2\cosh{c}} - A_0\right)\nonumber\\
&&\quad\quad\quad\quad\quad \quad\quad + \frac{\Lambda\nu}{ \sqrt{R}} +  \Gamma \nu^2 \bigg], \label{eq55a}\\
&&\dot{\gamma}\sin{2\theta} = -\gamma_2\frac{\kappa \mu R\sinh{c}}{2 \cosh^2{c}}\left( \frac{\mu R}{2\cosh{c}} - A_0\right)\!,\quad\quad\label{eq55b}\\
&&\cos{2\theta} = -\tanh{c}.  \label{eq55c}
\end{eqnarray}
\end{subequations}
Eq.~\eqref{eq55c} has a solution with $c\leq 0$ defined on $\frac{-\pi}{4} < \theta \leq \frac{\pi}{4}$, and another with $ c > 0$ on $\frac{\pi}{4} < \theta < \frac{3\pi}{4}$. Then, we can write $\sin{2\theta} = \sech{c}$, which transforms Eqs.~\eqref{eq55a} and \eqref{eq55b} into
\begin{subequations} \label{eq56}\begin{equation}
\begin{aligned}
\dot{\gamma} R\tanh c &= \frac{\gamma_1}{2} \bigg[\kappa\mu\left(\frac{\mu R}{2\cosh c}-A_0\right)\\
& \quad + \left(\frac{\Lambda\nu}{\sqrt{R}}+\Gamma\nu^2\right)\cosh c\bigg] ,\label{eq56a}
\end{aligned}
\end{equation}
\begin{equation}
\dot{\gamma} =-\frac{\gamma_2\kappa\mu R \tanh c}{2}  \left(\frac{\mu R}{2\cosh c}-A_0\right) \!. \label{eq56b}
\end{equation}
\end{subequations}

\subsection{Imperfect pitchfork bifurcation}
Let us find out how shear modifies the pitchfork bifurcation of Fig.~\ref{fig7}. For small shear, we expect an imperfect bifurcation (cf.~e.g.~Ref.~\cite{ios90}). For $\dot{\gamma}=0$, Eq.~\eqref{eq56b} has the solutions $c=0$ and $\cosh c= \mu R/(2A_0)$. Inserting the latter expression into Eq.~\eqref{eq56a}, we obtain the outer bifurcation diagram on the $(c,\Lambda)$ plane given by
\begin{eqnarray}
\Lambda=-\Gamma\nu\sqrt{R}=
-\Gamma\nu\sqrt{\frac{2A_0}{\mu}\cosh c}\,\,,\quad \mbox{and }\quad c=0. \label{eq57}
\end{eqnarray}
Expanding this expression around $c=0$, we obtain to leading order
\begin{subequations}\label{eq58}
\begin{eqnarray}
&&R= \frac{2A_0}{\mu}\cosh c\Longrightarrow R= \frac{2A_0}{\mu}\left(1+\frac{c^2}{2}\right)\!,  \label{eq58a}\\
&&\Lambda=-\Gamma\nu\sqrt{\frac{2A_0}{\mu}}\left(1+\frac{c^2}{4}\right)\!\Longrightarrow 
\frac{\Lambda-\Lambda^*}{\Lambda^*}=\frac{c^2}{4}. \quad\label{eq58b}
\end{eqnarray}\end{subequations}
Thus, the inner limit of the outer pitchfork bifurcation diagram is:
\begin{eqnarray}
 \left(\frac{\Lambda-\Lambda^*}{-\Lambda^*}+\frac{c^2}{4}\right)c=0. \label{eq59}
\end{eqnarray}
See Ref.~\cite{ben78} for background on matched asymptotic expansions. Note that all terms in Eq.~\eqref{eq59} are  $\mathcal{O}(c^3)$.

\begin{figure}[!b]
	\centering
    \includegraphics[width=\linewidth]{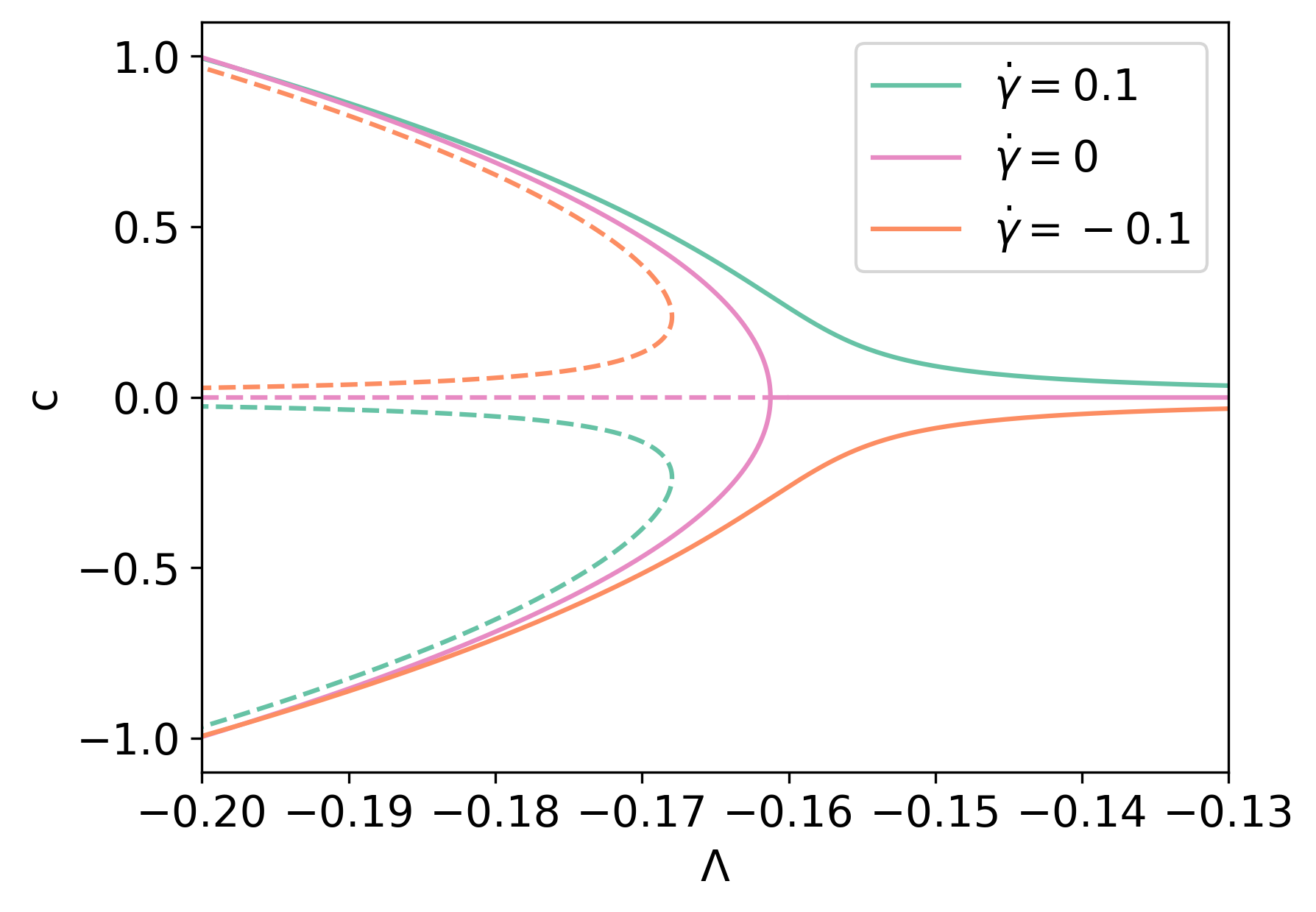}
    \caption{Imperfect pitchfork bifurcation of Eq.~\eqref{eq63a} appearing in the presence of a shear flow on top of the pitchfork bifurcation from figure \ref{fig7}. Green/orange curves correspond to positive/negative shear flow, respectively. Continuous (stable) and dashed (unstable) curves follow from linear stability with eigenvalues given by Eqs.~\eqref{eq70} and \eqref{eq71} below. Parameter values are $\kappa=\nu=\gamma_1=\gamma_2=1$, $A_0=\pi$, $\Gamma=0.1$, $\mu= 2.4166$, $\dot{\gamma}=\pm 0.1$.} 
    \label{fig11}
\end{figure}

To obtain the bifurcation diagram modified by a small $\dot{\gamma}$, we first substitute Eq.~\eqref{eq56b} into \eqref{eq56a}, thereby obtaining
\begin{eqnarray}
\left(\frac{\Lambda\nu}{\sqrt{R}}+\Gamma\nu^2\right)\!\sinh c =2\dot{\gamma}\left(\frac{1}{\gamma_2R}+ \frac{R \tanh^2c}{\gamma_1}\right)\!.\, \label{eq60}
\end{eqnarray}
Let us now consider how a small shear modifies the immediate neighborhood of the bifurcation point in Fig.~\ref{fig7}, 
\begin{equation}
R=R^*+r, \quad \Lambda=\Lambda^*+\lambda,\quad c=c,  \label{eq61}
\end{equation}
where $R^*=2A_0/\mu$
and
$r$, $c$, and $\lambda$ are small. Equation~\eqref{eq59} indicates that $\lambda=\mathcal{O}(c^2)$ and Eq.~\eqref{eq60} that $\dot{\gamma}=\mathcal{O}(c^3)$. Inserting Eq.~\eqref{eq61} into \eqref{eq60} and keeping only leading order terms, we obtain
\begin{equation}
r=- \frac{2\sqrt{R^*}}{\Gamma\nu} \,\lambda.  \label{eq62}
\end{equation}
We now substitute this expression into Eq.~\eqref{eq56b} and keep only leading order terms. The result is
\begin{subequations}\label{eq63}
\begin{eqnarray}
&&\!\left(\frac{\Lambda-\Lambda^*}{\Gamma\nu\sqrt{R^*}} +\frac{c^2}{4}\right) c =\dot{\gamma}_0, \label{eq63a}\\
&&\dot{\gamma}_0=\frac{\dot{\gamma}}{2\gamma_2\kappa A_0^2}. \label{eq63b}
\end{eqnarray}\end{subequations}
Hence the right hand side of Eq.~\eqref{eq63a}, given by $\dot{\gamma}_0$, acts as a small imperfection on the pitchfork bifurcation diagram, as shown in Figure \ref{fig11}. There we have depicted the imperfect bifurcation for both $\dot{\gamma}>0$ and its mirror image for $\dot{\gamma}<0$.

For positive $\lambda\gg 1$, $c\ll 1$ and Eq.~\eqref{eq63a} becomes the hyperbola 
\begin{equation}
\lambda\, c= \dot{\gamma}_0\Gamma\nu\sqrt{R^*},  \label{eq64}
\end{equation}
which is on the half plane having $\text{sign } c=\text{sign }\dot{\gamma}$. The other half plane contains a turning point having $d\lambda/dc=0$; see Fig.~\ref{fig11}. Together with Eqs.~\eqref{eq63}, this gives the turning point $(\lambda_0,c_0)$ with
\begin{subequations}\label{eq65}
    \begin{eqnarray}
&&c_0=(-2\dot{\gamma}_0)^\frac{1}{3}, \quad r_0= - \frac{2\sqrt{R^*}\lambda_0}{\Gamma\nu}, \label{eq65a}\\
&&\lambda_0\!=-3\Gamma\nu\sqrt{R^*}\!\left(\frac{\dot{\gamma}_0}{4}\right)^{2/3}. \label{eq65b} 
\end{eqnarray}
\end{subequations}

For large values of $|c|$, Eq.~\eqref{eq56b} yields Eq.~\eqref{eq58a}, $R=2A_0\cosh c/\mu$. Substituting this into Eq.~\eqref{eq60} and approximating $\tanh c=\pm 1$ for large $c$, we find
\begin{subequations}\label{eq66}
\begin{eqnarray}
\Lambda=-\sqrt{\frac{2A_0}{\mu}}\!\left(\Gamma\nu\mp\frac{4\dot{\gamma}A_0}{\gamma_1\mu\nu}\right)\! \sqrt{\cosh c}, \label{eq66a}
\end{eqnarray}
For consistency with the imperfect bifurcation diagram, we should have 
\begin{equation}
    \frac{4 \dot{\gamma}A_0}{\gamma_1\mu\Gamma\nu^2}<1.\label{eq66b}
\end{equation}
\end{subequations}

Eq.~\eqref{eq63} approximates Eq.~\eqref{eq60} and matches the outer solution for $\dot{\gamma}$ given by Eq.~\eqref{eq57}. In fact, the inner approximation of Eq.~\eqref{eq57} is Eq.~\eqref{eq59}, which clearly matches Eq.~\eqref{eq63} as $\dot{\gamma}\to 0$.  To find an uniformly valid bifurcation diagram at leading order, we add inner and outer solutions and subtract their common part. After multiplication by $\sqrt{R^*}$, the result is
\begin{eqnarray}
\!\left(\frac{\Lambda}{\Gamma\nu\sqrt{R^*}}+\sqrt{\cosh c}\right) c=\dot{\gamma}_0, \label{eq67}
\end{eqnarray}
where $\dot{\gamma}_0$ is given by Eq.~\eqref{eq63b}.

\subsection{Linear stability}

To find out the linear stability of the stationary solutions
-- this time also accounting for spatial degrees of freedom --
we linearize Eqs.~\eqref{eq50} around the stationary solutions obtained from the bifurcation equation \eqref{eq63a} with two simplifying assumptions: (i) we consider periodic boundary conditions on a rectangular box, and (ii) we assume $R-R^*=\mathcal{O}(c^2)$, $\Lambda-\Lambda^*=\mathcal{O}(c^2)$, $\dot{\gamma}=\mathcal{O}(c^3)$. The linearized equations contain terms that depend on $y$ due to the material derivatives. We can eliminate them by shifting $x\to x-\dot{\gamma}yt$. Then, if the unknowns in the linearized equations are proportional to $\exp[\sigma t+ik_1(x-\dot{\gamma}yt)+ik_2y]$, $\sigma$ are the eigenvalues of the matrix $A_{ij}$, where
\begin{subequations}\label{eq68}
    \begin{eqnarray}
 &&       A_{11}= \frac{\dot{\gamma}\sinh c}{\cosh^2c}-\gamma_1\!\left(\frac{\partial^2f}{\partial R^2}\!+\frac{K_Gk^2}{2}(1+\tanh^2\!c)\right)\!, \\ 
 &&        A_{12}=\frac{\dot{\gamma}R}{\cosh^3c}-\gamma_1\!\left(\frac{\partial^2f}{\partial R\partial c}+\frac{K_Gk^2R\sinh c}{2\cosh^3c}\right)\!,\\
  &&       A_{21}=-\gamma_2\!\left(\frac{\partial^2f}{\partial R\partial c}+\frac{K_Gk^2R\sinh c}{2\cosh^3c}\right)\!,\\
  &&       A_{22}=-\gamma_2\!\left(\frac{\partial^2f}{\partial c^2}+\frac{K_GR^2k^2}{2\cosh^4c}\right)\!,\\ 
  &&      A_{13}= -2\dot{\gamma}R\tanh^2c,\, A_{23}= -2\dot{\gamma}\tanh c,   \\ 
   &&     A_{31}\!=0,\, A_{32}\!=\! \frac{\dot{\gamma}}{\sinh(2c)},\, A_{33}\!=\!-\frac{\dot{\gamma}}{\sinh\! c}\!-\gamma_3K_Gk^2\!,\quad\quad
    \end{eqnarray}
\end{subequations}
and $k=\sqrt{k_1^2+k_2^2}$. Here $f=f_\text{sc}$ of Eq.~\eqref{eq33} and $R$, $c$, and $\theta$ are the stationary solutions given by Eqs.~\eqref{eq61}-\eqref{eq63}. Expanding around the bifurcation point $R^*$, $c^*=0$, $\Lambda^*$ with the scaling (ii), we obtain an approximation including up to $\mathcal{O}(c^2)$ terms.
\begin{subequations}\label{eq69}
    \begin{eqnarray}
 &&       A_{11}= -\gamma_1\!\left(\frac{\partial^2f}{\partial R^2}+\frac{K_Gk^2}{2}(1+c^2)\right)\!,\label{eq69a}\\ 
 &&        A_{12}=-\gamma_1\!\left(\frac{\partial^2f}{\partial R\partial c}+\frac{K_Gk^2R^* c}{2}\right)\!,\label{eq69b}\\
  &&       A_{21}=-\gamma_2\!\left(\frac{\partial^2f}{\partial R\partial c}+\frac{K_Gk^2R^* c}{2}\right)\!,\label{eq69c}\\
  &&       A_{22}\!=\!-\gamma_2\!\left(\frac{\partial^2f}{\partial c^2}+\frac{K_GR^*k^2\!}{2}(R^*+r-2R^*c^2)\!\right)\!,\quad\quad \label{eq69d}\\ 
  &&      A_{13}= A_{23}= A_{31}\!=0,\label{eq69e}\\
  && A_{32}\!=\! \frac{\dot{\gamma}}{2c},\quad A_{33}\!=\!-\frac{\dot{\gamma}}{c}-\gamma_3K_Gk^2,\label{eq69f}\\
  &&\frac{\partial^2f}{\partial R^2}= \frac{\kappa\mu^2(1-c^2)+\Gamma\nu^2}{4}-\frac{3\Gamma\nu^2r}{8R^{*2}}-\frac{\lambda\nu}{4R^{*3/2}},\label{eq69g}\\
  &&\frac{\partial^2f}{\partial c^2}=\frac{3\kappa A_0^2c^2}{2}, \quad\frac{\partial^2f}{\partial R\partial c}=-\kappa\mu A_0c. \label{eq69h}
    \end{eqnarray}
\end{subequations}
Expanding by minors, Eqs.~\eqref{eq69e} and \eqref{eq69f} imply that the determinant det$(A-\sigma I)$ is $(A_{33}-\sigma)$ times det$(A_{ij}-\sigma\delta_{ij})$ ($i,j=1,2$). Thus, one eigenvalue is
\begin{eqnarray}
    \sigma_3=-\frac{\dot{\gamma}}{c}-\gamma_3K_Gk^2.\label{eq70}
\end{eqnarray}
Eq.~\eqref{eq70} states that the stationary solutions with $\dot{\gamma}c>0$ (corresponding to continuous branches existing for all values of $\Lambda$
in Fig.~\ref{fig11}) are always stable. In contrast, the branches issuing from the turning points in Fig.~\ref{fig11} are unstable for wavenumbers on the interval $0<k^2<-\dot{\gamma}/(\gamma_3K_Gc)$. They will produce spatially non-homogeneous solutions when the size of the tissue exceeds a critical value. Here the kinetic coefficient $\gamma_3$ plays a stabilizing role.

The other two eigenvalues are those of the submatrix given by Eqs.~\eqref{eq69a}-\eqref{eq69d}. We find 
\begin{subequations}\label{eq71}
\begin{eqnarray}
&&\sigma_1= -\frac{\gamma_1}{4}\!\left(\kappa\mu^2+\frac{\Gamma\nu^2}{R^*}+2K_Gk^2\right)\! + \mathcal{O}(c^2),\label{eq71a}\\
&&\sigma_2\!=\!-\frac{\gamma_2}{2}\!\left[K_GR^*k^2(R^*\!+\!2r\!+\!2c^2)\!+\!\kappa A_0(3A_0c^2-\mu r)\right.\,\, \nonumber\\
&&\left.+ \frac{8\gamma_1A_0^2c^2(\kappa\mu-K_Gk^2/\mu)^2}{\gamma_1\!(\kappa\mu^2\!+\!\frac{\Gamma\nu^2}{R^*}\!+\!2K_Gk^2)\!-\!2\gamma_2\!K_GR^{*2}k^2}\right]\!\!+\! \mathcal{O}(c^3), \label{eq71b}
\end{eqnarray}
\end{subequations}
provided the denominator in the last term of Eq.~\eqref{eq71b} is $\mathcal{O}(1)$. If this denominator vanishes for some value of the parameters, the eigenvalues become 
\begin{eqnarray}
    \sigma_{1,2} \!&=&\!-\frac{\gamma_1}{4}\!\left(K_Gk^2+\frac{\kappa\mu^2}{2}+\frac{\Gamma\nu^2}{2R^*}\right)\! - \frac{2\gamma_2}{\mu^2}A_0^2K_Gk^2\quad\nonumber\\
   &\pm &\! \sqrt{\gamma_1\gamma_2} A_0|c| \left|\frac{K_Gk^2}{\mu}- \frac{\kappa\mu}{2}\right| + \mathcal{O}(c^2). \label{eq72}
\end{eqnarray}
In all cases, these eigenvalues are negative and the stability of the homogeneous phases under shear is decided by Eq.~\eqref{eq70} alone.

\section{Conclusions}\label{sec:9}
In this work, we have derived macroscopic hydrodynamic equations to describe a monolayer of confluent cells. We started from the mesoscopic  vertex model, which provides a convenient average single-cell free energy in terms of the cell area and perimeter. These quantities can be related to an average shape tensor, which can be written in terms of three fields: $R$ (trace, proportional to perimeter square), $c$ (anisotropy), and $\theta$ (angle of the director field describing the nematic-like alignment of elongated cells). Hydrodynamic equations follow from a coarse-graining procedure using Poisson brackets \cite{her21}. The reactive part of the hydrodynamic equations is a straightforward consequence of averaging the Poisson brackets of microscopic quantities, i.e.~density, momentum and shape tensor. The dissipative part of the equations depends on the choice of the kinetic coefficients and the average free energy. Our consistent choice produces a gradient system for homogeneous phases, which have the same dynamic stability properties as the thermodynamic stability of their homogeneous counterpart. 

Thus, we could recover the solid-liquid transition for a critical value of the line tension, which appears as a supercritical pitchfork bifurcation between isotropic cells ($c=0$) and elongated cells with nonzero anisotropy field $c$. Furthermore, we have analyzed how a homogeneous shear flow converts this transition into an imperfect pitchfork bifurcation. There, the continuous branches are stable, even including spatial degrees of freedom,
while the saddle-node branches 
are unstable vs.~spatially inhomogeneous states in sufficiently large monolayers.

In the future, the coarse-grained equations for a tissue obtained here can be investigated for 
several biologically relevant situations -- for instance, having spatially varying parameters in the equations. 
An interesting related question is the stability of a boundary between two
tissues that have different line tensions, as in antagonistic migration assays of two cell populations \cite{moi19,bon20}.
Conceptually, the most important next step would be to include 
activity into the coarse-graining process. Active, e.g. contractile, elements 
have been included in phenomenological and mechanical approaches to tissues \cite{tlili15,ish17}, 
and active dynamics similar to Eqs.~(\ref{eq3}), (\ref{eq4}) have been coarse-grained to yield Vicsek-type models \cite{bert06,Farr12}, However, a systematic coarse-graining of these effects on the tissue level have not yet been undertaken.\\

\acknowledgements
This work has been supported by the FEDER/Ministerio de Ciencia, Innovaci\'on y Universidades -- Agencia Estatal de Investigaci\'on grant PID2020-112796RB-C22, by the Madrid Government (Comunidad de Madrid-Spain) under the Multiannual Agreement with UC3M in the line of Excellence of University Professors (EPUC3M23), and in the context of the V PRICIT (Regional Programme of Research and Technological Innovation).

\end{document}